\documentclass{aa}  

\usepackage{graphicx}
\usepackage{txfonts}
\usepackage{siunitx,subfig,xspace,natbib}
\usepackage{hyperref}
\usepackage{amsmath}%
\usepackage{amsfonts}%
\usepackage{amssymb}%
\usepackage{mathrsfs}
\usepackage{arydshln}
\usepackage{cprotect}

\newcommand{\mrm}[1]{\mathrm{#1}}
\newcommand{\nuc}[2]{$\mrm{^{#2}#1}$}

\begin{document}

   \title{Nucleosynthesis constraints through $\gamma$-ray line measurements from classical novae}

   \subtitle{A hierarchical model for the ejecta of \nuc{Na}{22} and \nuc{Be}{7}}

\author{
  Thomas Siegert \and
  Sohan Ghosh \and
  Kalp Mathur \and
  Ethan Spraggon \and
  Akshay Yeddanapudi
}
\institute{
	Center for Astrophysics and Space Sciences, University of California, San Diego, 9500 Gilman Dr, 92093-0424, La Jolla, USA
  	\label{inst:ucsd}\\
  	\email{tsiegert@ucsd.edu}
}

   \date{Received January 7, 2021; accepted MM DD, YYYY}

\abstract
{Classical novae are among the most frequent transient events in the Milky Way, and key agents of ongoing nucleosynthesis.
Despite their large numbers, they have never been observed in soft $\gamma$-ray emission.
Measurements of their $\gamma$-ray signatures would provide both, insights on explosion mechanism as well as nucleosynthesis products.}
{Our goal is to constrain the ejecta masses of \nuc{Be}{7} and \nuc{Na}{22} from classical novae through their $\gamma$-ray line emissions at 478 and 1275\,keV. }
{We extract posterior distributions on the line fluxes from archival data of the INTEGRAL/SPI spectrometer telescope.
We then use a Bayesian hierarchical model to link individual objects and diffuse emission and infer ejecta masses from the whole population of classical novae in the Galaxy.}
{Individual novae are too dim to be detectable in soft $\gamma$-rays, and the upper bounds on their flux and ejecta mass uncertainties cover several orders of magnitude.
Within the framework of our hierarchical model, we can, nevertheless, infer tight upper bounds on the \nuc{Na}{22} ejecta masses, given all uncertainties from individual objects as well as diffuse emission, of $<2.0 \times 10^{-7}\,\mrm{M_{\odot}}$ (99.85th percentile).}
{In the context of ONe nucleosynthesis, the \nuc{Na}{22} bounds are consistent with theoretical expectations, and exclude that most ONe novae happen on white dwarfs with masses around $1.35\,\mrm{M_{\odot}}$.
The upper bounds from \nuc{Be}{7} are uninformative.
From the combined ejecta mass estimate of \nuc{Na}{22} and its $\beta^+$-decay, we infer a positron production rate of $<5.5 \times 10^{42}\,\mrm{e^+\,s^{-1}}$, which would make at most 10\,\% of the total annihilation rate in the Milky Way.} 
 

   \maketitle
%

\section{Introduction}\label{sec:intro}
Classical novae (CNe) are thermonuclear explosions on the surface of white dwarfs (WDs) and among the most frequent transient phenomena within a galaxy.
Despite their high occurrence rate in the Milky Way of $50 \pm 25\,\mrm{yr^{-1}}$ \citep{Shafter2017_novarate}, only about 20\,\% are seen in UVOIR wavelengths during one year.
At X-ray \citep{Ness2007_Swiftnovae} and GeV \citep{Franckowiak2018_FermiLATnovae} energies, only a few tens of objects in total could be analysed in detail to date.
At MeV energies, CNe have never been observed, leaving a large gap in the understanding of nucleosynthesis in these events, as well as in the transition region between thermal and non-thermal processes.

The absolute nucleosynthesis yields from CNe are very uncertain and have only been inferred indirectly from abundance ratios in the expanding nova clouds, the latter of which generally agree very well with theoretical expectations \citep[e.g.,][]{Livio1994_nova_abundances}.
Using $\gamma$-rays, the interiors of these objects can be studied and absolute abundances determined \citep[e.g.,][]{Clayton1974_novae,Clayton1981_novae7Li,Leising1987_nova_511,Hernanz2006_novae,Hernanz2014_nova}.
Among the most promising messengers of ongoing nucleosynthesis in CNe are \nuc{Be}{7} and \nuc{Na}{22}, which emit soft $\gamma$-rays at 478\,keV and 1275\,keV, respectively, as by-products of the decays to their stable daughter-nuclei \nuc{Li}{7} and \nuc{Ne}{22}.

Individual objects are too dim for the best current $\gamma$-ray line spectrometer telescope, INTEGRAL/SPI \citep{Winkler2003_INTEGRAL,Vedrenne2003_SPI}, to be detected beyond a few hundred pc.
But since many objects contribute to the total Galactic luminosity, we attempt to constrain the ejecta masses of \nuc{Be}{7} and \nuc{Na}{22} from the whole population of known and unknown CNe.
The cumulative effect of this population of `sub-threshold' sources is expected to result in a radioactive glow along the Galactic plane in both isotopes.
This effect is well-measured in the case of \nuc{Al}{26}, from the nucleosynthesis within massive stars \citep[e.g.,][]{Diehl2006_26Al,Bouchet2015_26Al,Pleintinger2019_26Al}.
In the case of CNe, earlier measurements with COMPTEL focused on the diffuse emission of the 1275\,keV line.
In the inner Galaxy ($|l| < 30^{\circ}$), \citet{Jean2001_ONenovae1275} found in an upper limit of the \nuc{Na}{22} ejecta mass from the diffuse emission of ONe novae (see Sec.\,\ref{sec:nova_review}) of $<3 \times 10^{-7}\,\mrm{M_{\odot}}$.
The strongest constraints so far for an individual object was found by \citet{Iyudin1999_NovaCygni1992}, with a limit of $<2.1 \times 10^{-8}\,\mrm{M_{\odot}}$ from Nova Cygni 1992.
A similar, but less smooth extended emission can also be expected from the 478\,keV line, which however has not been considered yet.
The most stringent upper bound on the ejecta mass of \nuc{Be}{7} has been found by \citet{Siegert2018_V5668Sgr} for the CO nova V5668 Sgr, with $<1.2 \times 10^{-8}\,\mrm{M_{\odot}}$, assuming a distance of 1.6\,kpc.
All these limits are close to the highest ejecta mass estimates, for example from UV measurements \citep[e.g.,][finding $<0.7 \times 10^{-8}\,\mrm{M_{\odot}}$ for V5668 Sgr]{Molaro2016_V5668}, which however rely on a canonical value for the total ejecta mass, being itself uncertain by one order of magnitude.

In addition to the diffuse part, more than $100$ individual CNe are known to have occurred during the time of the INTEGRAL observations since 2003, hence their contribution must also be taken into account.
This is possible in the framework of a Bayesian hierarchical model \citep{Gelman2013_BDA3} in which we attempt to link the physics of individual CNe and hence determine ejecta mass estimates from the whole CN population for both CO and ONe novae.

This paper is structured as follows:
In Sec.\,\ref{sec:nova_theory}, we introduce the nucleosynthesis that is expected to occur in CNe, and predict individual and cumulative $\gamma$-ray signals from these expectations in general.
Sec.\,\ref{sec:data_sets} describes the INTEGRAL/SPI data set, followed by Sec.\,\ref{sec:nova_catalogue} which includes the sample of known CNe during the time of the observations.
The general data analysis to extract fluxes from the raw count data is shown in Sec.\,\ref{sec:data_analysis}, together with the refined approach of the Bayesian hierarchical model.
We present our results from individual sources, diffuse emission, and hierarchical modelling in Sec.\,\ref{sec:results}.
In Sec.\,\ref{sec:discussion}, we discuss these results in terms of CN nucleosynthesis constraints and the contribution of ONe novae to the Galactic positron puzzle.
We summarise and conclude in Sec.\,\ref{sec:conclusion}.

\section{Nuclear Astrophysics Expectations for Novae}\label{sec:nova_theory}

\subsection{Explosive Burning}\label{sec:nova_review}
Nova explosions are the result of mass accretion onto a WD in a close binary system.
The explosion itself is described as a thermonuclear runaway reaction of accreted material that gradually becomes degenerate, heats up under the additional pressure from still-accreting matter, and finally ignites when specific conditions are met \citep[e.g.,][]{Jose1998_novae}.
Depending on the composition, and consequently mass of the WD, different isotopes, up to mass number $\sim 40$, are produced and ejected into its surroundings \citep[e.g.,][]{Jose1998_novae,Jose2001_novae_intermediate,abcdefgh,Starrfield2020_COnovae}.
Additionally, the composition of the accreted material has to be considered as well as its fraction in the final mixture on the WD's surface, resulting in a broad range of possible abundance ratios from this explosive nucleosynthesis.
In particular, two types of generic WD compositions are typically considered: CO WDs in the mass range $0.6$--$1.15\,\mrm{M_{\odot}}$ and ONe WDs in the mass range above $1\,\mrm{M_{\odot}}$.
Thus, from an independent measurement of a WD's mass, the composition and CN type can roughly be determined \citep[e.g.,][]{Gil-Pons2003_novae}.

In hydrodynamical models, typical accretion rates are on the order of $10^{-10}\,\mrm{M_{\odot}\,yr^{-1}}$, which results in an envelope mass of about $M^{env} \sim 10^{-5}\,\mrm{M_{\odot}}$ during an accretion time of $10^5\,\mrm{yr}$.
The explosion sets in at a temperature around $2 \times 10^7\,\mrm{K}$ and reaches a peak of several $10^{8}\,\mrm{K}$ about a minute after the ignition.
The complete thermonuclear runway results in the canonical total ejecta mass of $M_{tot}^{ej} \sim 10^{-5}\,\mrm{M_{\odot}}$.
In a scenario where WDs gradually gain mass and eventually explode as type Ia supernovae, it would be expected that $M^{env} > M^{ej}_{tot}$ \citep{Starrfield2020_COnovae}.
Expansion velocities are typically expected in the range of $500$--$3000\,\mrm{km\,s^{-1}}$.

UVOIR observations of CNe typically happen around maximum light when the objects are detected days to weeks after the initial explosion, as the expanding nova cloud is opaque of low-energy photons \citep{Gomez-Gomar1998_novae}.
Insight about the onset of the explosion is therefore linked to strong assumptions from theoretical modelling.
During the explosive nucleosynthesis, also short-lived isotopes, such as \nuc{N}{13} ($\tau_{13} = 10\,\mrm{min}$) and \nuc{F}{18} ($\tau_{18} = 110\,\mrm{min}$), are produced, which decay by positron emission.
The positrons would quickly find electrons and lead to a strong 511\,keV line during the first hour after the explosion.
This however has never been observed, mainly because this strong annihilation flash happens days to weeks before the CN is detected.
Also retrospective searches for individual known objects or complete archives have not found any of these counterparts \citep{Skinner2008_nova511_retrospective}.

An alternative to complete the information in CN explosions is provided by longer-lived nucleosynthesis products.
In particular, the electron-capture decay of \nuc{Be}{7} ($\tau_7 = 76.78\,\mrm{d}$) results in an excited state of \nuc{Li}{7} which de-excites quickly by the emission of a $\gamma$-ray photon at 478\,keV, so that enough \nuc{Be}{7} is still present for follow-up observations after the detection of a CN.
The synthesis of \nuc{Be}{7} is believed to occur via $\mrm{^3He(\alpha,\gamma)^7Be}$ in both CN types.
The CO nova abundances of \nuc{Be}{7}, however, are expected to be about one order of magnitude larger than for ONe novae \citep{Jose1998_novae}.
In contrast, ONe novae are expected to produce major amounts of \nuc{Na}{22} and \nuc{Al}{26}, whereas those abundances in CO are expected to be several orders of magnitude smaller.
The production of \nuc{Na}{22} happens mainly through proton captures on seed nuclei, with the main reactions $\mrm{^{20}Ne(p,\gamma)^{21}Na(p,\gamma)^{22}Mg(\beta^+)^{22}Na}$ and $\mrm{^{20}Ne(p,\gamma)^{21}Na(\beta^+)^{21}Ne(p,\gamma)^{22}Na}$ as part of the nuclear reaction network in the NeNaMgAl region.
With a life time of $\tau_{22} = 3.75\,\mrm{yr}$, \nuc{Na}{22} decays (90\.\% $\beta^+$; 10\,\% electron capture) to \nuc{Ne^*}{22} which emits a $\gamma$-ray photon at 1275\,keV.
This allows us to search for \nuc{Na}{22} even years after the explosion, and from multiple sources at the same time (see Sec.\,\ref{sec:diffuse_emission}).
The long lifetime of \nuc{Al}{26} ($\tau_{26} = 1.05\,\mrm{Myr}$) results in a mixture with other, more dominant \nuc{Al}{26} sources in the Galaxy, so that individual CN explosions cannot be traced back any more, and only the cumulative diffuse emission remains.

We therefore restrict this study to the intermediate lifetime isotopes \nuc{Be}{7} and \nuc{Na}{22}.
Models predict \nuc{Be}{7} yields between $10^{-11}$ and $10^{-8}\,\mrm{M_\odot}$ for a single CO nova event \citep[e.g.,][]{Jose2001_novae_intermediate,Jose2001_novaegamma,Jose2006_novae,Hernanz2006_novae,abcdefgh,Hernanz2014_nova,Starrfield2020_COnovae}, obtaining an upper limit for the detectability of 500\,pc with INTEGRAL/SPI in the most optimistic case.
The yields for \nuc{Na}{22} in ONe novae might reach up to $10^{-6}$--$10^{-7}\,\mrm{M_\odot}$, depending on the WD mass, composition, and mixture \cite{abcdefgh}.
However, older estimates also suggest lower \nuc{Na}{22} ejecta masses in ONe novae with up to $10^{-8}\,\mrm{M_\odot}$ \citep[e.g.,][]{Jose1998_novae}.

During INTEGRAL's now 18 mission years, about $100$ CNe are known to have happened inside the Milky Way (see Sec.\,\ref{sec:nova_catalogue}), whereas about $900$ are expected owing to the CN rate.
Because the occurrence rate is so large compared to the decay rates of \nuc{Be}{7} and \nuc{Na}{22}, an equilibrium mass of radioactive material inside the whole Milky Way can be expected, which supersedes that of a single CN event.
In the following, we detail out predictions from first principles what to expect from $\gamma$-ray measurements.

\subsection{Gamma-Rays from Individual Objects}\label{sec:individual_objects}
From the ejected mass $M_a^{ej}$ of each isotope $a$, the maximum $\gamma$-ray flux $F_{0,a}$ at the time of the explosion $T_0$ from a CN at distance $d$ can be estimated by
\begin{equation}
F_{0,a} = \frac{1}{4 \pi d^2}\frac{p_a^{\gamma} M_a^{ej}}{m_a \tau_a}\mrm{,}
\label{eq:max_flux}
\end{equation}
\noindent where $m_a$ is the atomic mass of isotope $a$, $\tau_a$ its characteristic lifetime, and $p_a^{\gamma}$ the probability of the daughter nucleus to emit a photon due to nuclear de-excitation \citep{Thielemann2018_AwR}.
From the radioactive decay law, the flux as a function of time $t$ follows an exponential decay,
\begin{equation}
F_a(t) = F_{0,a} \exp(-(t-T_0)/\tau_a)\Theta(t-T_0)\mrm{,}
\end{equation}
\noindent where $\Theta(t-T_0)$ is the Heaviside function, setting starting time (explosion date) to $T_0$.
Given the angular resolution of SPI of $\sim 2.7^{\circ}$, individual CNe appear as point sources.
The full spatio-temporal model thus reads
\begin{equation}
F_a(l,b,t) = F_a(t) \delta(l-l_0) \delta(b-b_0)\mrm{,}
\end{equation}
\noindent where $(l_0,b_0)$ are the coordinates of a CN in Galactic longitude and latitude, respectively.
For the two considered isotopes in this study, \nuc{Be}{7} and \nuc{Na}{22}, the atomic masses are $m_{7} = 7.017\,\mrm{u}$ and $m_{22} = 21.994\,\mrm{u}$, their decay times are $\tau_7 = 76.8\,\mrm{d}$ and $\tau_{22} = 3.75\,\mrm{yr}$, and the probabilities to emit a 478\,keV and 1275\,keV photon, respectively, are $p_7^{\gamma} = 0.1044$ and $p_{22}^{\gamma} = 0.999$.
The maximum flux at 478\,keV and 1275\,keV is therefore
\begin{equation}
F_{0,7} = 22887 \left( \frac{M_7^{ej}}{\mrm{M_{\odot}}} \right) \left( \frac{d}{\mrm{kpc}} \right)^{-2}\,\mrm{ph\,cm^{-2}\,s^{-1}}
\label{eq:flux_ps_7}
\end{equation}
\noindent and
\begin{equation}
F_{0,22} = 3838 \left( \frac{M_{22}^{ej}}{\mrm{M_{\odot}}} \right) \left( \frac{d}{\mrm{kpc}} \right)^{-2}\,\mrm{ph\,cm^{-2}\,s^{-1}}\mrm{.}
\label{eq:flux_ps_22}
\end{equation}
Clearly, individual objects can only be seen by SPI if the distance is of the order of a few hundreds of pc, given its nominal $3\sigma$ narrow line sensitivity\footnote{\url{https://www.cosmos.esa.int/web/integral/observation-time-estimator}} of $7 \times 10^{-5}\,\mrm{ph\,cm^{-2}\,s^{-1}}$ (478\,keV) and $5 \times 10^{-5}\,\mrm{ph\,cm^{-2}\,s^{-1}}$ (1275\,keV).
We note that, especially for \nuc{Be}{7}, a sensitivity estimate or distance threshold that is based on the maximum flux becomes flawed rather quickly because the exponential decay law decreases the number of received photon faster than the significance will increase by a typical square-root of exposure time scaling.
Likewise, for \nuc{Na}{22} the Doppler broadening of several $1000\,\mrm{km\,s^{-1}}$ adds significantly to the instrumental resolution, so that the actual CN line sensitivities are worse (cf. Sec.\,\ref{sec:results}).
Nevertheless, the additional information of the exponential decay can be used when applying the SPI response (Sec.\,\ref{sec:general_SPI_analysis}) to this model to predict the number of expected counts, and thus provide a more robust estimate of the ejected mass.

\subsection{Gamma-Rays from Diffuse Emission}\label{sec:diffuse_emission}

On average, only about 20\,\% of the expected CNe per year in the Milky Way are detected in UVOIR wavelengths (see Sec.\,\ref{sec:nova_catalogue}).
These and all undetected sources still contribute to the Galactic-wide $\gamma$-ray emission.
Since the decay times of both \nuc{Be}{7} and \nuc{Na}{22} are longer than the average waiting time between two CNe, $\tau_N := 1/R_N \approx 7\,\mrm{d}$, the Galaxy in 478 and 1275\,keV can be described by a diffuse glow of unseen CNe due to the radioactive build-up.
The quasi-persistent $\gamma$-ray luminosity of the population of CNe in a galaxy can be estimated by the sum over all unknown individual objects $n$, hence
\begin{equation}
	L_a^{diff} = \sum_{n} L_{a,n}^{PS} = \sum_n \frac{M_a p_a^{\gamma}}{\tau_a m_a} e^{-\frac{t-T_{0,n}}{\tau_a}} \Theta(t-T_{0,n}) = p_a^{\gamma} \frac{M_a^{ej}}{m_a} R_N\mrm{.}
	\label{eq:diff_lumi}
\end{equation}
In Eq.\,(\ref{eq:diff_lumi}), $L_{a,n}^{PS}$ is the luminosity of a point source $n$, emitting photons from isotope $a$, and $T_{0,n} = n/R_N = n \tau_N$ is the average explosion time of each object.
We note that the final expression is independent of the decay time $\tau_a$ for any isotope.
This is reasonable since Eq.\,(\ref{eq:diff_lumi}) only describes the average luminosity of an entire galaxy that is producing an average mass of $M_a^{ej}$ at a rate $R_N$.
If instead the diffuse (population) luminosity is considered as the sum of individual CN luminosities with maximum $L_{0,a}^{PS}$, Eq.\,(\ref{eq:diff_lumi}) becomes
\begin{equation}
L_a^{diff} = L_{0,a} R_N \tau_a\mrm{,}
\end{equation}
which defines whether the galactic-wide luminosity is dominated by a single source ($\tau_N \gg \tau_a$) or the population ($\tau_N \ll \tau_a$).
Examples would be \nuc{Al}{26} mainly from massive stars and their supernovae with a decay time of $\tau_{26} = 1.05\,\mrm{Myr}$ compared to the core-collapse supernova rate in the Milky Way of $R_{CCSN} = 0.02\,\mrm{yr^{-1}}$ \citep[e.g.][]{Diehl2006_26Al}, showing that on the order of $10^4$ supernovae contribute to the diffuse 1.8\,MeV emission.
Conversely, \nuc{Ti}{44}, which is also produced mainly in core-collapse supernovae, has decay time of only $86\,\mrm{yr}$, so that the Milky Way in \nuc{Ti}{44} decay photons is dominated by one (or a few) supernova remnants at each time \citep{The2006_44Ti}.

The diffuse $\gamma$-ray flux can be estimated similarly, resulting in a direct conversion between a measured flux and the luminosity, or the ejected mass of each object, given a known CN rate:
\begin{equation}
F_{a}^{diff} = \sum_n \frac{L_a^{diff}}{4 \pi d_n^2} = \omega L_a^{diff}\mrm{,}
\label{eq:flux_diff_inf_sum}
\end{equation}
where $\omega = \sum_n (4 \pi d_n^2)^{-1}$ is related to the `effective distance' of the diffuse emission, taking into account the probability of a CN to occur at a distance $d_n$ from the Sun.
The conversion factor $\omega$ can be determined in two equivalent ways by assuming a 3D density distribution of the population of CNe, which we will describe in the following.
\citet{Shafter2017_novarate} showed that CNe in the Milky Way can be described as a linear combination of a De Vaucouleurs profile ($\rho_1(x,y,z)$) for the bulge and a doubly exponential disk ($\rho_2(x,y,z)$), where the weights of the components are taken as the relative CN rates in bulge and disk, $f_1 = 0.1$ and $f_2 = 0.9$.
The normalised density profiles, such that $\int\,dV\,\rho(x,y,z) = 1$, are given by
\begin{equation}
	\rho_1(x,y,z) = \frac{16 a^{17/2}}{2027025 \pi^{3/2} R_e^3} \exp\left(-a\left(\frac{R}{R_e}\right)^{1/4}\right)\left(\frac{R}{R_e}\right)^{-7/8}
	\label{eq:devauc}
\end{equation}
and
\begin{equation}
	\rho_2(x,y,z) = \frac{1}{4 \pi r_e^2 z_e} \exp\left(-\frac{r}{r_e}\right) \exp\left(-\frac{|z|}{z_e}\right)\mrm{,}
	\label{eq:exp_disk}
\end{equation}
with $R = \sqrt{x^2 + y^2 + z^2}$, $a = -7.669$, $R_e = 2.7\,\mrm{kpc}$, $r = \sqrt{x^2 + y^2}$, $r_e = 3.0\,\mrm{kpc}$, and $z_e = 0.25\,\mrm{kpc}$.
\begin{figure}[!hbtp]
	\centering
	\includegraphics[width=1.0\columnwidth,trim=0.0in 0.1in 0.0in 0.6in,clip=true]{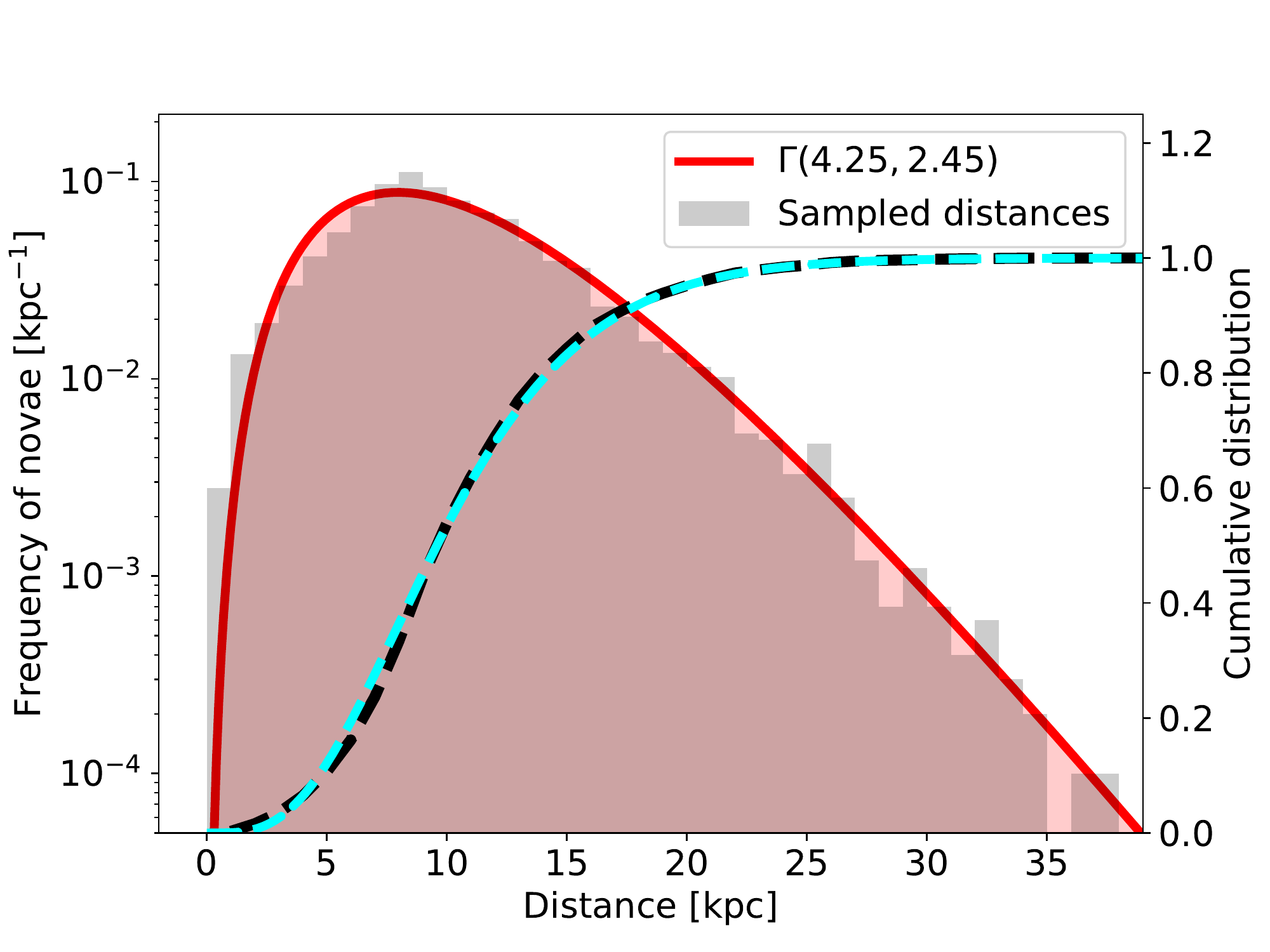}
	\caption{Distance distribution of CNe according to Eqs.\,(\ref{eq:devauc}) and (\ref{eq:exp_disk}) via rejection sampling. Shown are $10^4$ samples (gray histogram, corresponding to a time scale of 200\,yr, left axis), and an approximation with the $\Gamma$-distribution in red. The cumulative distributions are shown in black for the sample and in cyan for the $\Gamma$-distribution (right axis).}
	\label{fig:distance_samples}
\end{figure}
The distribution of distances can now be directly sampled from the density profiles, for example via 3D rejection sampling of $(x,y,z)$-coordinates, from which the distances are calculated as $d = \sqrt{(x-x_s)^2+(y-y_s)^2+(z-z_s)^2}$, where $(x_s,y_s,z_s) = (8.179,0,0.020)$\,kpc is the position of the Sun.
We show $10^4$ samples of distances calculated from the distribution $\rho_{tot} = f_1\rho_1 + f_2\rho_2$ in Fig.\,\ref{fig:distance_samples}.
We find that this distribution can adequately approximated by a $\Gamma$-distribution with $\alpha_d = 4.25$ and $\beta_d = 1/2.45$, with the expectation value $\alpha_d/\beta_a = 10.4$\,kpc.
We will use this distribution of CN distances later in Sec.\,\ref{sec:unknown_distances} to define a prior for objects with unknown distances.
With a large sample size, the infinite sum in Eq.\,(\ref{eq:flux_diff_inf_sum}) converges to $\omega = 0.00291\,\mrm{kpc^{-2}}$, or an `effective distance' of $d_{eff} = (L^{diff}/(4 \pi F^{diff}))^{1/2} = 5.23\,\mrm{kpc}$.

\begin{figure}[!htbp]
	\centering
	\includegraphics[width=1.0\columnwidth,trim=0.5in 1.0in 0.8in 2.1in,clip=true]{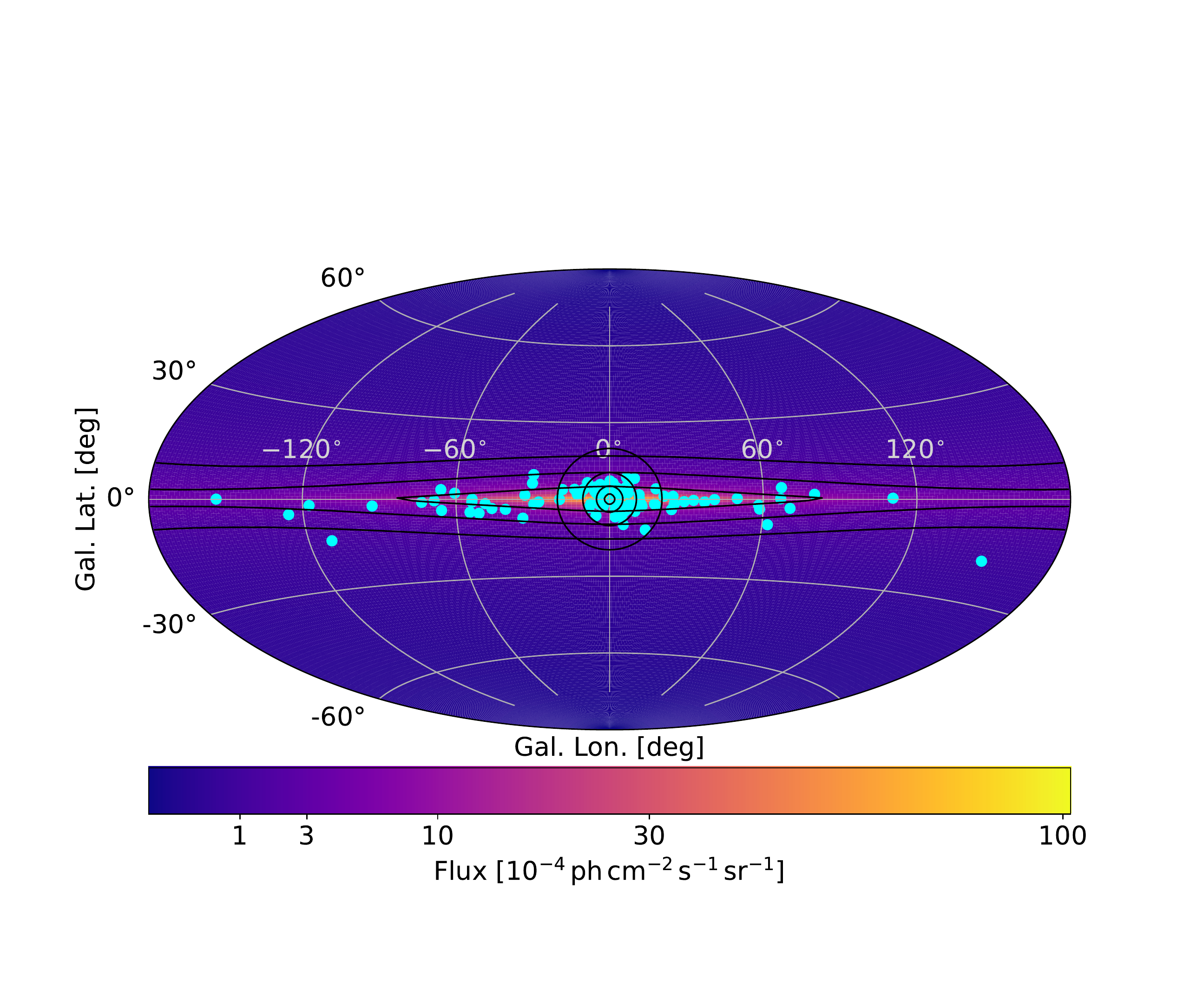}
	\caption{Diffuse emission template from line-of-sight-integrated nova density distribution, Eq.\,(\ref{eq:template_map_astro}), together with nova sample (cyan points). The contours indicate the De Vaucouleurs profile for the bulge and the exponential disk.}
	\label{fig:diffuse_map_plus_sources}
\end{figure}

Alternative to estimating the infinite sum, we calculate the diffuse emission map exactly by line-of-sight integration,
\begin{equation}
	F^{diff}(l,b) = \frac{1}{4 \pi} \int_0^{+\infty}\,ds\,\rho(x'(s),y'(s),z'(s))\mrm{,}
\end{equation}
where $(x'(s),y'(s),z'(s)) = (x_s-s\cos(l)\cos(b), y_s-s\sin(l)\cos(b)), z_s-s\sin(b))$ is the line-of-sight vector starting from the Sun in all directions $(l,b)$ \citep[c.f. also][]{Erwin2015_IMFIT}.
The total integrated flux of this map is consequently
\begin{equation}
	F^{diff} = \int\,d\Omega F^{diff}(l,b)\mrm{.}
\end{equation}
The intrinsic luminosity of a given density distribution is given by
\begin{equation}
	L^{diff} = \int\,d\Omega \int_0^{+\infty}\,ds\,s^2\,\rho(x'(s),y'(s),z'(s))\mrm{,}
\end{equation}
so that the conversion factor reads
\begin{equation}
	\omega = \frac{1}{4 \pi}\frac{\int\,d\Omega\,\int_0^{+\infty}\,ds\,\rho(x'(s),y'(s),z'(s))}{\int\,d\Omega \int_0^{+\infty}\,ds\,s^2\,\rho(x'(s),y'(s),z'(s))}\mrm{.}
\end{equation}
Diffuse quasi-persistent $\gamma$-ray emission from the population of CNe in the Milky Way is therefore
\begin{equation}
	F_a^{diff}(l,b) = L_a^{diff} \omega F^{diff}(l,b)\mrm{.}
	\label{eq:template_map_astro}
\end{equation}
In Fig.\,\ref{fig:diffuse_map_plus_sources}, we show the diffuse emission template and overlay all individual CNe that we considered for this study (Sec.\,\ref{sec:nova_catalogue}).
For the two isotopes, the diffuse flux can finally be related to the ejected mass by
\begin{equation}
	F_7^{diff} = 178 \left(\frac{M_7^{ej}}{\mrm{M_{\odot}}}\right) \left(\frac{R_N}{\mrm{yr^{-1}}}\right)\,\mrm{ph\,cm^{-2}\,s^{-1}}
	\label{eq:flux_diff_be7}
\end{equation}
and
\begin{equation}
F_{22}^{diff} = 528 \left(\frac{M_{22}^{ej}}{\mrm{M_{\odot}}}\right) \left(\frac{R_{ONe}}{\mrm{yr^{-1}}}\right)\,\mrm{ph\,cm^{-2}\,s^{-1}}\mrm{,}
\label{eq:flux_diff_na22}
\end{equation}
where $R_{ONe} \approx \frac{1}{3} R_N$ is the ONe nova rate \citep{Gil-Pons2003_novae}.
Using Eqs.\,(\ref{eq:flux_diff_be7}) and (\ref{eq:flux_diff_na22}) and the theoretical expectations, Sec.\,\ref{sec:nova_theory}, a diffuse flux for the 478\,keV line of the order of $10^{-8}$--$10^{-5}\,\mrm{ph\,cm^{-2}\,s^{-1}}$ can be expected.
In the energy region of the \nuc{Be}{7} decay line, there is strong Galactic background emission, mainly from the ortho-positronium continuum and Inverse Compton scattering \citep[e.g.,][]{Churazov2005_511,Churazov2011_511,Jean2006_511,Bouchet2010_511,Siegert2016_511,Siegert2019_lv511}, which we take into account for estimating the ejected \nuc{Be}{7} in a later step (Sec.\,\ref{sec:diffuse_emission_results}).
Even though the expected fluxes for individual objects as well as the diffuse emission of the 478\,keV line are barely scratching the sensitivity of SPI, a combined fit, taking into account that all objects share similar physics, can provide stronger limits on the ejected masses.
Likewise, the 1275\,keV line from \nuc{Na}{22} can be expected to show a diffuse flux of the order of $10^{-5}$--$10^{-4}\,\mrm{ph\,cm^{-2}\,s^{-1}}$.
This is well within the sensitivity threshold of SPI.
However, the instrumental background line at 1275\,keV does not follow strictly the variation of cosmic-ray intensity defined by the solar cycle, but builds up as a function of mission time and thus increases the background at these energies \citep{Diehl2018_BGRDB}.

\begin{figure*}[!ht]
	\centering
	\includegraphics[width=1.0\textwidth,trim=2.0in 0.2in 1.5in 0.6in,clip=true]{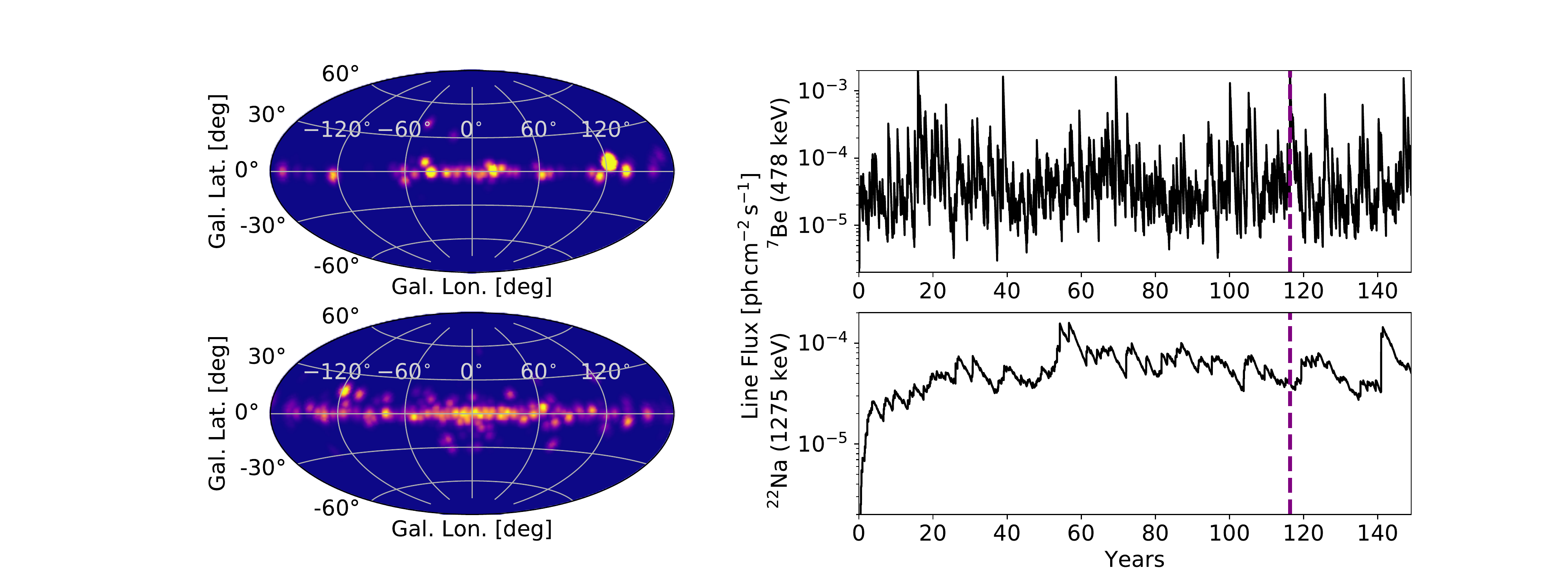}
	\caption{Population synthesis of nova events in a Milky Way like galaxy. \textit{Left}: Shown is year $116.3$ of the modelled Poisson process, where a nearby CO nova at $(l,b) = (+123^{\circ},+8^{\circ})$ outshines the remaining galaxy in 478\,keV emission (\textit{top}). At the same time, 1275\,keV emission is not enhanced (\textit{bottom}). The angular resolution is chosen according to SPI's $2.7^{\circ}$ (FWHM). \textit{Right}: Time profiles (light curves) of the 478\,keV and 1275\,keV since the beginning of the synthesis for a duration of $\sim 150\,\mrm{yr}$. This time scale is enough to reach convergence and to show characteristic features. The vertical purple line indicates the time shown in the plots on the left.}
	\label{fig:time_profile_lines}
\end{figure*}

\subsection{Expectations from Cumulative Signals}\label{sec:cumulative_expectations}
From above considerations and theoretical expectations, we compile a first-order population synthesis model for both isotopes, and compare this to the expected background of our chosen data set (see Sec.\,\ref{sec:data_sets}) to obtain a signal-to-noise ratio.
In both cases, we sample 3D-positions according to the combined 3D-density distributions, Eqs.\,(\ref{eq:devauc}) and (\ref{eq:exp_disk}), from which distances and Galactic coordinates are calculated.
The occurrence of CNe in the Milky Way is a Poisson process \citep{Ross2008_StochasticProcesses} with a rate of $R_N \approx 50\,\mrm{yr^{-1}}$ for CO novae and $\approx \frac{1}{3} R_N$ for ONe novae.
This means the waiting time for each CNe after the last one is exponentially distributed.
We therefore sample the waiting times as $\Delta T \sim \mrm{Exp}(R_N)$.
This provides the explosion date of event $n$ by $T_{0,n} = \sum_i^n \Delta T_i$.
For the logarithm of ejected masses of ONe novae, we assume a normal distribution $\lg M_{22}^{ej} \sim \mathscr{N}(-8.25,0.5^2)$ and for CO novae $\lg M_{7}^{ej} \sim \mathscr{N}(-10,1^2)$, presenting conservative estimates from theoretical expectations (Sec.\,\ref{sec:nova_review}).
The widths of the distributions are chosen to cover the plausible range of theoretical ejecta masses, and to not produce unreasonably large $\gamma$-ray fluxes which would have already been seen by previous instruments.
In a more realistic version of this population synthesis model, the ejecta masses would be distributed according to the mass distribution of WDs on which CNe happen.
Since both distributions are fairly uncertain, we stick to the general case and only include the range of theoretical models from \citet{abcdefgh} and \citet{Starrfield2020_COnovae}.
Finally, fluxes at the sampled positions are calculated via Eqs.\,(\ref{eq:flux_ps_7}) and (\ref{eq:flux_ps_22}).

We show the characteristic time profile for one realisation in both lines integrated across the whole sky in Fig.\,\ref{fig:time_profile_lines}.
Clearly, the flux of the 478\,keV line appears more erratic since the lifetime of \nuc{Be}{7} is shorter compared to \nuc{Na}{22}.
But since the lifetime of \nuc{Be}{7} is still larger than the average waiting time of about one week, at any given time, the total integrated flux averages to $2$--$11 \times 10^{-5}\,\mathrm{ph\,cm^{-2}\,s^{-1}}$.
Large peaks, such as around year 18 and 23, are due to individual objects that happened close to the observer.
The `diffuse' part of the 478\,keV emission can be estimated from this to be on the order of $10^{-5}\,\mathrm{ph\,cm^{-2}\,s^{-1}}$, whereas at each given time, the total flux is dominated by $\sim 10$ individual objects (cf. $R_N \tau_7 \approx 10.5$).
For \nuc{Na}{22}, the decreased ONe nova rate yields to a quasi-persistent flux between $4$ and $9 \times 10^{-5}\,\mathrm{ph\,cm^{-2}\,s^{-1}}$.
In the case of the 1275\,keV line, the diffuse flux is dominant, around $5 \times 10^{-5}\,\mathrm{ph\,cm^{-2}\,s^{-1}}$, and individual objects only contribute significantly if they are close to the observer, such as the event at year 23.

\begin{figure}[!th]
	\centering
	\includegraphics[width=1.0\columnwidth,trim=0.0in 0.4in 0.3in 0.2in,clip=true]{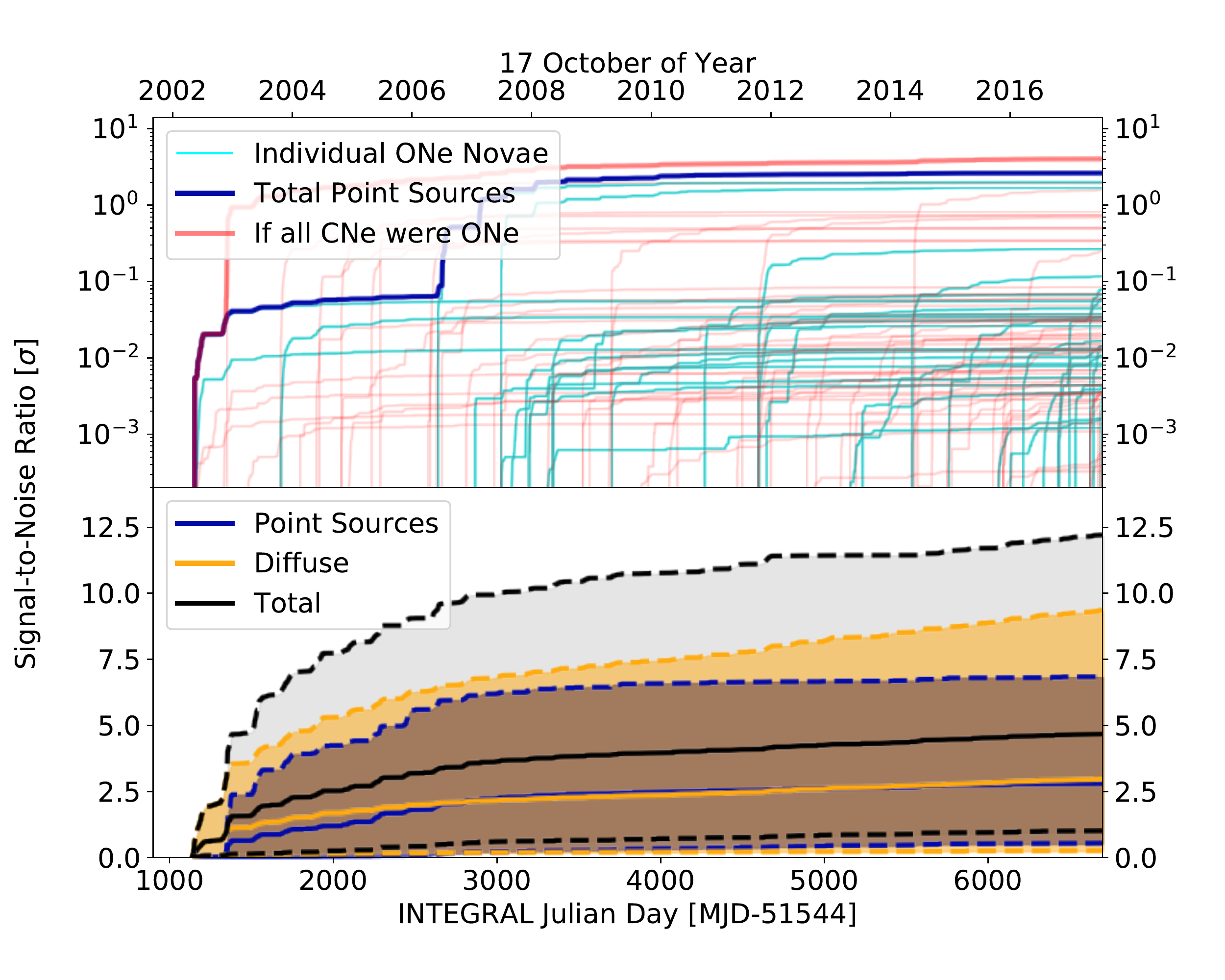}
	\caption{Estimated signal-to-noise ratios given the SPI data set (Sec.\,\ref{sec:data_sets}) and the range of plausible \nuc{Na}{22} ejecta masses for our sample of 97 novae. \textit{Top}: The SNR of individual objects does not increase with $\sqrt{T_{exp}}$ since the exponential decay is faster than the accumulation of 1275\,keV photons (cyan lines). The cumulative signal of point sources saturates on a time scale of a few years (dark blue). If all objects are assumed to be ONe novae, the SNR would be about three times higher (pale red). \textit{Bottom}: Comparison between cumulative point source emission significance and diffuse emission significance during the INTEGRAL mission time scale. The bands contain the 68th percentile from 1000 realisations of the population synthesis model (boundaries marked by dashed lines, median by solid lines; Sec.\,\ref{sec:cumulative_expectations}), to distribute ejecta masses among sources and diffuse emission.}
	\label{fig:SNR_estimates}
\end{figure}

We estimate the range of possible signal-to-noise ratios (SNRs), given these assumptions, using a generic background model (Sec.\,\ref{sec:data_analysis}), and 1000 realisations of our CN sample (Sec.\,\ref{sec:nova_catalogue}) and the diffuse emission (Eq.\,(\ref{eq:template_map_astro})) in Fig.\,\ref{fig:SNR_estimates} for the case of \nuc{Na}{22}.
In both cases, individual objects are, most of the time, below the detection threshold of SPI.
However the cumulative signal of 97 targets that are known to emit at these wavelengths, result in an average SNR of $\sim 5\sigma$ for \nuc{Be}{7} and $\sim 3\sigma$ for \nuc{Na}{22}.
In the diffuse emission case, the 478\,keV line would be seen on average with a SNR of $\sim 1\sigma$, and the 1275\,keV line with $\sim 3\sigma$.
Therefore, both emission lines would show a SNR of $\sim 5\sigma$ when considering this INTEGRAL data set.
However, the allowed range of theoretical ejecta masses leads to considerable uncertainties, being consistent with undetectable signals for SPI even after 16 years.

\section{SPI Data Sets}\label{sec:data_sets}

We use publicly available SPI data between 2003 and 2018, including INTEGRAL revolutions 43--1951.
Apart from a few exceptions, such as V5668 Sgr \citep{Siegert2018_V5668Sgr}, INTEGRAL did not purposely target CN outbursts, so that the exposure times of our selected sample (see Sec.\,\ref{sec:nova_catalogue}) often contain large gaps or objects are only observed in the partially coded field of view.
In a first step, we apply selection criteria based on orbital parameters and instrumental count rates and sensors:
Until the year 2015, we select orbital phases 0.1--0.9 to avoid the Van Allen radiation belts.
Afterwards, due to an orbit manoeuver, we only select orbital phases 0.15--0.85.
We exclude times when the count rate of the SPI anticoincidence shield is increased significantly, for example due to solar flares.
We further exclude times when the cooling plate temperatures show a difference larger than 1\,K.
Revolutions 1554--1558 are removed from the data set due to the outburst of the microquasar V404 Cygni, being the brightest soft $\gamma$-ray source in the sky during this time.
Revolutions in which a detector failure happened are completely excluded as well.

In a second step, we fit this data set with our instrumental background model and the diffuse emission map (Sec.\,\ref{sec:data_analysis}), and remove all pointed observations that show residuals larger than $10\sigma$.
This removes about $0.1\,\%$ of all selected pointings.
We show the resulting exposure map in Fig.\,\ref{fig:expo_map}.

\subsection{\nuc{Be}{7}}\label{sec:be7_data_set}
The electron capture of \nuc{Be}{7} results in a photon at 477.62\,keV from an excited state of \nuc{Li}{7} ($p_7^{\gamma} = 0.1044$).
We use all SPI `single events', i.e. photons that only interact once with the detector, resulting in $100105$ pointed observations, with a typical exposure time between 1800 and 3600\,s.
The total dead-time corrected exposure time, taking into account failed detectors, is 180.1\,Ms.

The spectral resolution at 478\,keV is about 2.1\,keV \citep{Diehl2018_BGRDB}.
Additional astrophysical Doppler broadening of the line can be expected from the expansion velocity of the nova ejecta, typically in the range of $500$--$3000\,\mrm{km\,s^{-1}}$ (see Sec.\,\ref{sec:nova_review}).
To include more than 99\,\% of the $\gamma$-ray line photons in a single bin for an ejecta velocity of $2000\,\mrm{km\,s^{-1}}$, we use an 8\,keV wide energy bin between 474 and 482\,keV.
This includes 7.0\,keV broadening due to homologous ejecta, and is added in quadrature to the instrumental resolution.

\subsection{\nuc{Na}{22}}\label{sec:na22_data_set}
The decay of \nuc{Na}{22} to \nuc{Ne}{22}, either via $\beta^+$-decay ($p_{22}^+ = 0.904$) or electron capture, results in a $\gamma$-ray photon of 1274.58\,keV with $p_{22}^{\gamma} = 0.999$.
At this energy range, `electronic noise' features in the SPI spectrum can emerge, which can be filtered through pulse shape discrimination (PSD).
The selection for PSD events reduces the size of the data set to $99880$ pointings, and the exposure time to 179.6\,Ms.
We note that the fluxes resulting from PSD events have to be re-scaled by an efficiency correction factor $\approx 1/0.85$.

SPI's spectral resolution around 1275\,keV is about 2.8\,keV, so that we perform our analysis in a single 20\,keV energy bin between 1265 and 1285\,keV to account for possible astrophysical Doppler broadening (as above, Sec.\,\ref{sec:be7_data_set}).

\begin{figure}[!htbp]
	\centering
	\includegraphics[width=1.0\columnwidth,trim=0.0in 1.0in 0.6in 2.0in,clip=true]{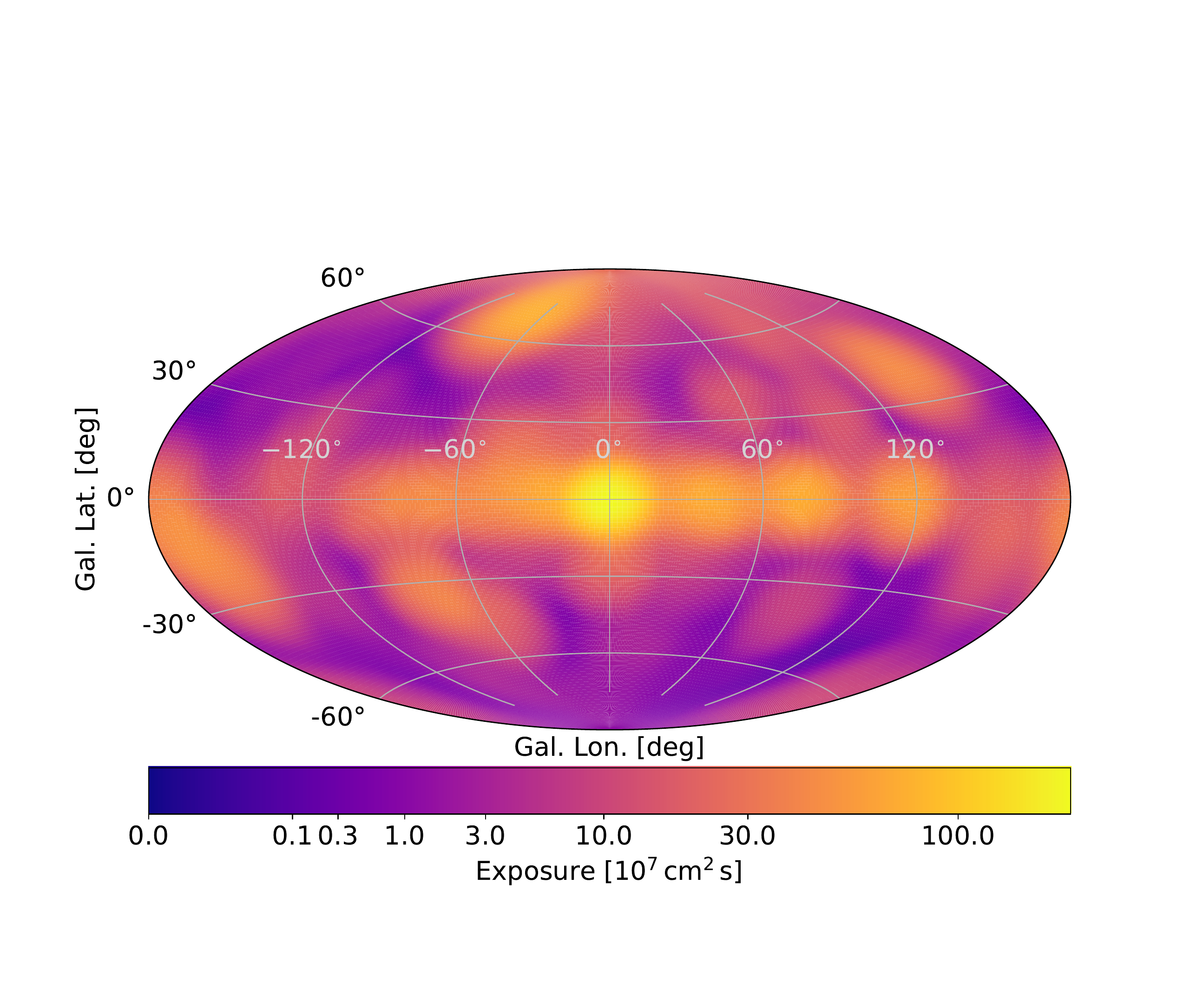}
	\caption{Exposure map of the selected data set. The effective areas around 478 and 1275\,keV are $\sim 80$ and $\sim 55\,\mrm{cm^{2}}$, respectively.}
	\label{fig:expo_map}
\end{figure}

\section{Nova Sample}\label{sec:nova_catalogue}

\subsection{Selecting from Databases and Catalogues}\label{sec:sample_selection}
In order to infer information about the ejecta masses from flux measurements, the distance to the objects and their unambiguous classification as CNe is important.
While many astrophysical transients are called `novae', we clearly want to include only the thermonuclear explosions on the surface of WDs, termed `classical novae'.
For current soft $\gamma$-ray instrumentation, any CN outside the Milky Way is hopeless to detect, so we restrict our selections to our own Galaxy, i.e. objects with galacto-centric distances of $\lesssim 25\,\mrm{kpc}$ (cf. Fig.\,\ref{fig:distance_samples}).
This is justified considering the expected quasi-persistent 478 and 1275\,keV line fluxes from M31 or the Large Magellanic Cloud (LMC):
Using the same formalism from Sec.\,\ref{sec:cumulative_expectations} with adjusted CN rates ($R_{N}(\mrm{M31}) = 100\,\mrm{yr^{-1}}$, $R_{N}(\mrm{LMC}) = 2\,\mrm{yr^{-1}}$) and distances ($d(\mrm{M31}) = 780\,\mrm{kpc}$, $d(\mrm{LMC}) = 50\,\mrm{kpc}$), we expect at most $10^{-7}\,\mrm{ph\,cm^{-2}\,s^{-1}}$ (1275\,keV) and $9 \times 10^{-9}\,\mrm{ph\,cm^{-2}\,s^{-1}}$ (478\,keV) from the LMC and $2 \times 10^{-8}\,\mrm{ph\,cm^{-2}\,s^{-1}}$ (1275\,keV) and $2 \times 10^{-9}\,\mrm{ph\,cm^{-2}\,s^{-1}}$ (478\,keV) from M31.

There are several comprehensive and well-maintained CN catalogues available online, however scattered into multiple publications and with different information.
We first select CNe from recent peer-reviewed articles, in particular \citet{Ozdonmez2018_novacat}, \citet{Shafter2017_novarate}, \citet[][SMEI]{Hounsell2016_SMEInovae}, \citet[][VVV]{Saito2013_VVVnovae}, and \citet[][SMARTS]{Walter2012_SMARTSnovae}.
We complement the information by individual websites that host and update CN lists.
These include Koji Mukai's `List of Recent Galactic Novae' (2008--2020)\footnote{\url{https://asd.gsfc.nasa.gov/Koji.Mukai/novae/novae.html}}, the CBAT `List of Novae in the Milky Way' (17th century -- 2010)\footnote{\url{http://www.cbat.eps.harvard.edu/nova_list.html}}, Bill Gray's `List of Galactic Novae' (`Project Pluto'; 17th century -- 2019)\footnote{\url{https://www.projectpluto.com/galnovae/galnovae.htm}}, Christian Buil's `Nova Corner' (1999--2015)\footnote{\url{http://www.astrosurf.com/buil/us/spe7/novae.htm}}, as well as the ARAS Spectral Data Base of Novae (2012--2020)\footnote{\url{http://www.astrosurf.com/aras/Aras_DataBase/Novae.htm}}.

INTEGRAL launched on Oct 17th 2002 \citep{Winkler2003_INTEGRAL} and publicly accessible data is available starting with INTEGRAL revolution 43 ($\mrm{MJD} = 52683$; Feb 13th 2003).
Owing to the lifetime of \nuc{Na}{22} of 3.75\,yr, it is also reasonable to include CNe before the launch and first measurements of  INTEGRAL.
Given the shorter lifetime of \nuc{Be}{7} of 0.21\,yr and to make a coherent sample, we include all objects from Jan 1st 2002 ($\mrm{MJD} = 52275$) to Jun 30th 2018 ($\mrm{MJD} = 58299$), i.e. $16.5\,\mrm{yr}$, and those which were at least in the partially coded field of view of SPI ($\sim 30^{\circ}$ corner to corner) for 100\,ks.

We include all objects whose type is either a `fast', `moderately fast', `slow', or otherwise termed CN.
This also excludes nova-named objects, such as V838 Mon \citep{Bond2003_V838Mon} and similarly classified events.
The absolute magnitude in UVOIR wavelengths is no selection criterion for this study.
We do not include known recurrent CNe in our sample as they would require additional special treatment considering their long-term light curves and ejecta distribution.
This further excludes in particular U Sco (last two known outbursts in 1999 and 2010), RS Oph (1985 and 2006), and IM Nor (1920 and 2002).

In total, this provides a sample of `known' objects within the considered time frame of $97$ CNe.
Given the CN rate of $\sim 50\,\mrm{yr^{-1}}$, about $\sim 12\,\%$ of all expected objects are included as point sources with known positions in this study (Sec.\,\ref{sec:individual_objects}).
The remaining $\sim 88\,\%$ of `unknown' sources, plus hundreds of CNe from before the start of the considered time frame, then make up the diffuse component of the $\gamma$-ray emission (Sec.\,\ref{sec:diffuse_emission}).
The full list of objects can be found in Tab.\,\ref{tab:results_table}.

\subsection{Handling Unknown Distances}\label{sec:unknown_distances}
For $44$ CNe in this sample -- mainly older objects --, a distance estimate is available from the different input catalogues and websites.
If two or more estimates are available for one object, we used the most recent value.
Most of the distance estimates for CNe come from the maximum magnitude relation with decline time \citep[MMRD,][]{Zwicky1936_MMRD,McLaughlin1940_MMRD,Buscombe1955_MMRD}.
For objects located inside a certain cluster or galaxy, this method provides reliable estimates for the population distance.
However, individual CNe inside the Milky Way still carry uncertainties on their distances on the order of 30--50\,\%.
For example, the distance estimate of $3 \pm 1$\,kpc  for V5115 Sgr (Tab.\,\ref{tab:results_table}) would suggest an ejecta mass uncertainty of at least $66\,\%$ ($\approx 0.2\,\mrm{dex}$), ignoring the flux uncertainties.

The problem of converting estimated fluxes to ejecta masses for objects with unknown distance is even more severe:
Considering the distance distribution of CNe inside the Milky Way from the point of Earth, Fig.\,\ref{fig:distance_samples}, it is clear that most sources are expected around the Galactic centre.
However, an object found toward the Galactic anticentre with unknown distance will most certainly not be at such a large distance - albeit still possible, but only with a probability $P(7 < d < 9) \approx 17\,\%$.
Very UVOIR-bright CNe could be located close to Earth, however with unavailable distance estimate:
The probability that any object is located inside a sphere of radius 0.5\,kpc (1.0\,kpc) is $P(d \leq 0.5) \approx 3 \times 10^{-5}$ ($5 \times 10^{-4}$).
Given the CN rate, less than one object is expected within these distances during the INTEGRAL observations, consistent with no $\gamma$-ray detection so far.
These considerations make the distribution in Fig.\,\ref{fig:distance_samples} and its approximation with $d_{unknown} \sim \Gamma(\alpha_d = 4.25,1/\beta_d = 1/2.45)$ a reasonable first-order distance estimator for any CNe inside the Milky Way, independent of its direction as seen from Earth.
We will use the approximated distribution as prior information for inference in Sec.\,\ref{sec:no_pooling} to construct a full posterior for the ejecta masses, properly taking into account all uncertainties.

We estimate the impact of using this generic distribution by naive scaling factors as above:
As described in Sec.\,\ref{sec:diffuse_emission}, the expectation value of distances is $\langle d_{unknown} \rangle = 10.4$\,kpc, with a standard deviation $\sigma_{d_{unknown}} = 5.1$\,kpc.
Half of all CNe observed from Earth are found within a sphere of 9.6\,kpc, thus including the Galactic bulge, nearby high-latitude objects, as well as the Galactic anticentre, and thus also covering most `known' CN distances (Tab.\,\ref{tab:results_table}).
Considering the symmetric 90\,\% interval around the expectation value, a canonical distance uncertainty of $7.6$\,kpc can be given.
This makes the ejecta masses of objects with unknown distances uncertain by at least 150\,\% ($\approx 0.4$\,dex).
Given the large total volume of the Milky Way, this appears as a reasonable estimate.

\section{SPI Data Analysis}\label{sec:data_analysis}

\subsection{General Method}\label{sec:general_SPI_analysis}
SPI data $d_{jpe}$ are detector ($j$) triggers per unit time (pointing $p$) and energy ($e$), and consequently follow the Poisson distribution,
\begin{equation}
	d_{jpe} \sim \mrm{Poisson}(m_{jpe}) = \frac{m_{jpe}^{d_{jpe}}e^{-m_{jpe}}}{d_{jpe}!}\mrm{,}
	\label{eq:Poisson_dist}
\end{equation}
where $m_{jpe}$ is the (modelled) rate parameter.
In our analysis, the energy is fixed and we will omit the index $e$ in the following.
The likelihood of measuring $d_{jp}$ counts in the selected data set $D$, given a model $M$ with expectation $m_{jp}$ is
\begin{equation}
	\mathscr{L}(D|M) = \prod_{jp} \mrm{Poisson}(d_{jp}|m_{jp})\mrm{.}
\end{equation}
We model the SPI counts as a linear combination of sky components and background models,
\begin{equation}
	m_{jp} =  \sum_{i=1}^{N_N+1} F_{i,p} R_{jp}^{lb} S_{i,lb} + \sum_{k={L,C}} \beta_k B_{k,jp}\mrm{.}
	\label{eq:general_SPI_model}
\end{equation}
In Eq.\,(\ref{eq:general_SPI_model}), $N_N$ is the number of CNe in our sample, plus one model component to account for diffuse emission.
$F_{i,p}$ is the flux of sky model $S_{i,lb}$ to which the imaging response function $R_{jp}^{lb}$ is applied (mask coding).
The background model $B_{k,jp}$ is split into two components, $k=L$ and $C$, accounting for instrumental line and continuum background, respectively.
Temporal variations in the background are determined by the parameters $\beta_k$, scaling the amplitudes of the two components \citep[c.f.][]{Siegert2019_SPIBG}.
The expected flux of point-like sources (individual CNe) is determined according to Eqs.\,(\ref{eq:flux_ps_7}) and (\ref{eq:flux_ps_22}).
Since the exponential decay times are known, the pointing-to-pointing variation can be fixed in Eq.\,(\ref{eq:general_SPI_model}), and only one parameter is fitted for each source and isotope.
The expected diffuse flux is modelled according to Eqs.\,(\ref{eq:flux_diff_be7}) and (\ref{eq:flux_diff_na22}).
In the individual CN case, the spatial model is one point at the coordinates of the CN.
For diffuse emission, the flux is constant in time and distributed according to the line-of-sight-integrated density structure Eq.\,(\ref{eq:template_map_astro}).
Finally, the fit parameters of interest are the maximum flux $F$ at the time of explosion $T_0$, which can be split into a function of ejecta mass $M^{ej}$ and distance $d$ for point sources, and mass and CN rate $R_N$ for diffuse emission.

Clearly, this would result in an ill-defined likelihood function if the distances and the CN rate are not constrained.
In the case of a known distance for CNe in our sample, we incorporate a normal prior on the distance.
If the distance to an object is not known, we use the information that the CN happened `inside the Milky Way', which translates into a distance prior according to a $\Gamma$-distribution  $d_{unknown} \sim \Gamma(\alpha_d=4.25,\beta_d=1/2.45)$ (see Secs.\,\ref{sec:diffuse_emission} and \ref{sec:unknown_distances}).
This sets any object for which the distance is unknown on average to $\alpha_d/\beta_d = 10.4$\,kpc with a standard deviation $(\alpha_d/\beta_d^2)^{1/2} \approx 5.1$\,kpc (cf. Fig.\,\ref{fig:distance_samples}).
For the CN rate, we set a prior according to the most recent literature value in units of $\mrm{yr^{-1}}$ of $R_N \sim \mathscr{N}_+(50,25^2)$ (index $+$: truncated at zero) for all CN types and $R_{ONe} = \frac{1}{3} R_N$ for ONe novae \citep[][]{Shafter2017_novarate}.
Note that especially the prior for the CN rate is particularly broad, and can, in principle, be consistent with zero.
A zero rate, however, is unphysical since CNe do happen throughout the year constantly.
We nevertheless include this extreme to show a broad range for the remaining parameter space.
We set a uniform prior on the logarithm of the ejecta mass, $\lg M^{ej} \sim \mathscr{U}(-11,-4)$, so that each decade considered in the parameter space obtains the same prior probability, and a wide range of theoretical and otherwise plausible values are sampled.
The lower bound on the ejecta mass prior of $10^{-11}\,\mathrm{M_{\odot}}$ is data driven, as above considerations (Sec.\,\ref{sec:cumulative_expectations}) show that SPI is incapable to probe these values except for distances below 100\,pc.

Details on the background modelling procedure can be found in \citet{Diehl2018_BGRDB} and \citet{Siegert2019_SPIBG}.
In short, the background is modelled by long-term monitoring of the complete SPI spectrum and the spectral response changes.
This results in a background and response data base per INTEGRAL orbit and detector for several 100 background lines, and the underlying continuum.
Since the background dominates the measured count rate at any time, and in addition the coded mask pattern from celestial objects smears out on a time scale of tens of pointings (typically 50--100 in one INTEGRAL orbit), a background response can be constructed from this data base.
The short-term pointing-to-pointing variation is fixed by an onboard counting rate, in our case the saturating Germanium detector events.
Finally, any variation that is not captured by this procedure is handled by the introduction of the background re-scaling parameters $\beta_k$.
In both cases, we find that one parameter per INTEGRAL orbit per background component is enough to provide an adequate fit.
The total number of fitted parameters for one object finally is twice the number of INTEGRAL orbits, plus two for the flux, i.e. $n_{par} = 2 \cdot 1674 + 2 = 3350$.
For the 478\,keV line (1275\,keV), the number of data points excluding dead detectors is $1598964$ ($1595430$).

\subsection{Hierarchical Modelling Approach}\label{sec:hierarchical_models}
\begin{figure}[!htbp]
	\centering
	\includegraphics[width=1.0\columnwidth,trim=0.1in 0.2in 0.1in 0.2in,clip=true]{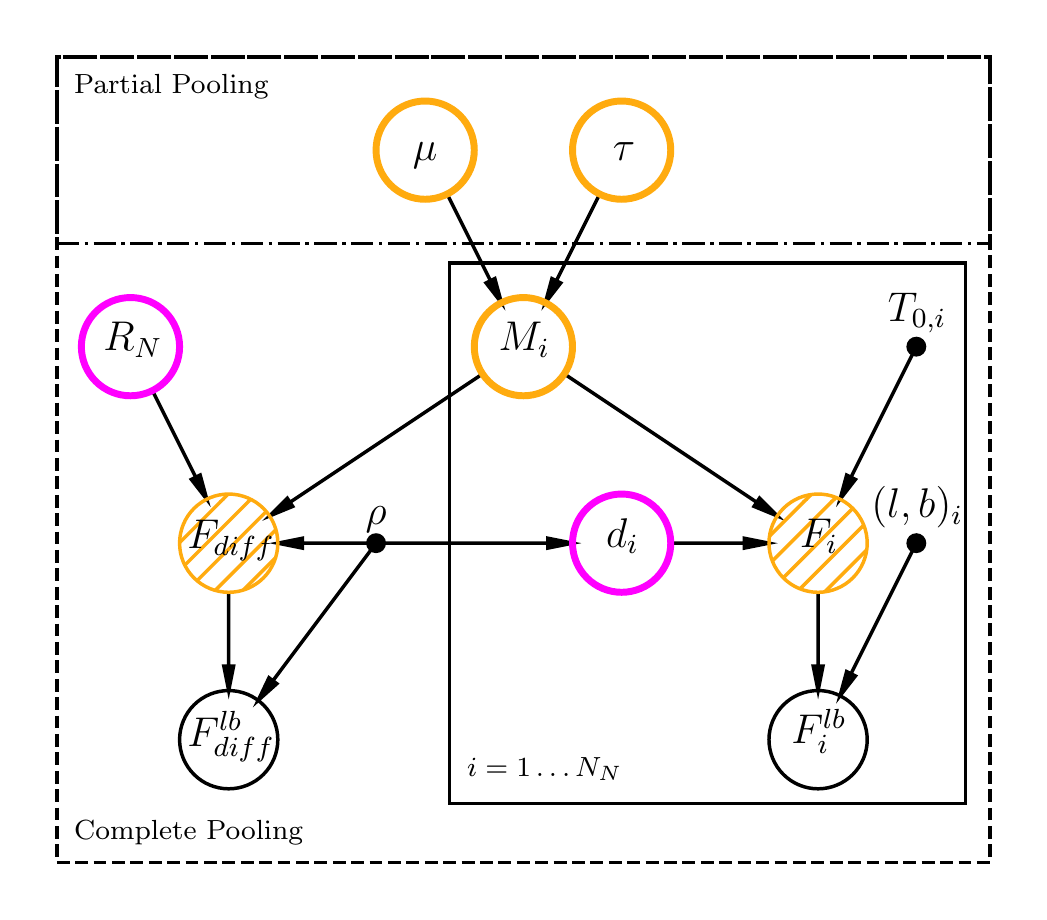}
	\caption{Complete graph of different model variants used in this study. Using the source models individually, i.e. being agnostic about a common underlying physical process, results in the most conservative estimates in a \textit{no pooling} setting (Sec.\,\ref{sec:no_pooling}). The opposite extreme assumes that all objects eject the same amount of matter, so that instead of $\sim 100$ different masses, only one is fitted. This resembles \textit{complete pooling} and defines tight constraints but is subject to bias if the true variance is large (Sec.\,\ref{sec:complete_pooling}). A compromise between no and complete pooling is achieved by invoking the hyper-parameters $\mu$ and $\tau$ that determine both the mean and the spread of ejecta masses, if any (\textit{partial pooling}, dash-dotted box, Sec.\,\ref{sec:partial_pooling}).}
	\label{fig:model_graphical}
\end{figure}

In order to combine the extracted posterior distributions of the flux, we apply a hierarchical model \citep{Gelman2013_BDA3} in three different steps.
In Fig.\,\ref{fig:model_graphical}, we show the full graphical model, as separated into `interesting' parameters (orange), nuisance parameters (magenta), fixed parameters (black dots), resulting spatio-temporal models (black circles), and their interdependencies (arrows).
Considering the sky emission from top to bottom, the hierarchy flows from superordinate hyper-parameters $\mu$ and $\tau$, which describe the distribution of ejecta masses $M_i$ of each CN $i$, which then determines the fluxes, given additional parameters:
Depending on the distances $d_i$, or in the diffuse case the CN rate $R_N$, absolute fluxes $F_i$ and $F_{diff}$ result via Eqs.\,(\ref{eq:flux_ps_7}\&\ref{eq:flux_ps_22}) and (\ref{eq:flux_diff_be7}\&\ref{eq:flux_diff_na22}).
These are either point-like at source positions $(l,b)_i$ or distributed according to the line-of-sight integration of $\rho$ (cf. Eqs.\,(\ref{eq:devauc}\&\ref{eq:exp_disk}) and (\ref{eq:template_map_astro})), resembling the sky models $F_{i}^{lb}$ and $F_{diff}^{lb}$, respectively.
For individual objects, the discovery times $T_{0,i}$ are known\footnote{The explosion times are expected 2--10 days before the discovery \citep{Gomez-Gomar1998_novae}.
This defines our `systematic uncertainties' in the case of CO novae of $\exp(0.1) \approx 10\,\%$ and ONe novae of $\exp(0.005) \approx 0.5\,\%$ -- much smaller than the statistical uncertainties.}, after which the flux is exponentially decaying according to isotope-specific parameters (Eq.\,\ref{eq:max_flux}).
The diffuse emission component is constant in time.
In particular, this scheme is included in our Poisson process, Sec.\,\ref{sec:cumulative_expectations} -- now we trying to infer the values of $M_i$ and/or $\mu$ and $\tau$.

Ideally, all sky models are convolved with the imaging response to convert them into a SPI-compatible format, and then fitted simultaneously to the raw SPI count data, together with the two-component background model.
Given the large number of data points, fitted parameters, and especially their interdependence in the full hierarchy (see Secs.\,\ref{sec:no_pooling}--\ref{sec:partial_pooling}), this is computationally very expensive and not feasible without major amounts of high-performance computing hours.
Instead, we apply the hierarchical model in a two step process:

First, we `extract' the fluxes of each of the $N_N = 97$ CNe and the diffuse emission individually (Eq.\,\ref{eq:general_SPI_model}).
This results in posterior distributions for the fluxes of all sky models considered.
To first order, these posteriors follow a $\Gamma$-distribution\footnote{In particular, the posterior of a Poisson rate is exactly $\Gamma$-distributed if the conjugate prior for the Poisson distribution is used, the $\Gamma$-distribution \citep{Gelman2013_BDA3}.}, so that instead of the previous $1.5$ million data points, we condense the information into the two shape parameters, $\alpha$ and $1/\beta$ of the $\Gamma$-distribution for each sky model.
Thus, in a second step, we apply the hierarchical model to a reduced data set consistent of $(97+1) \cdot 2$ shape parameters.

Since the actual measurement information (Poisson counts) is lost in this procedure, the resulting combined values are approximations to the true distribution.
We already note here that none of the sources considered show a significance above background of more than $2.5\sigma$, so that most flux posteriors are sharply peaked at zero flux, with a long tail according to the $\Gamma$-distribution.

\subsubsection{No Pooling}\label{sec:no_pooling}
The most agnostic view about the physics at play is provided by treating each object individually (\textit{No Pooling}).
This means we `let the data speak for themselves' if un- or weakly informative priors are used, as is the case for our general method:
the flux extraction step, Sec.\,\ref{sec:general_SPI_analysis}, is equivalent to a \textit{No Pooling} analysis.
Since we are interested in the ejecta masses, we fit the two shape parameter of the flux posteriors, in particular their expectation values $\langle F_i \rangle = \alpha_i/\beta_i$, according to the likelihood
\begin{equation}
	P(\langle F_i \rangle\, | \,\beta_i, M_i^{ej}, d_i, R_N) = \prod_{i=0}^{N_N+1} \Gamma(\langle F_i \rangle ; \tilde{F_i}\beta_i,\beta_i)\mrm{,}
	\label{eq:no_pooling_sampling}
\end{equation}
where $\tilde{F_i}$ are the latent fluxes according to Eqs.\,(\ref{eq:flux_ps_7},\ref{eq:flux_ps_22},\ref{eq:flux_diff_be7}\&\ref{eq:flux_diff_na22}).
This obtains -- by definition -- a distribution for the flux of each object consistent with the input distribution, and an expectation value $\langle F_i \rangle = \tilde{F_i}$.
By adding the distance (or CN rate) information in a prior yields the posterior distribution of ejecta masses for each individual object (and diffuse emission) independently, according to the measured flux and its asymmetric uncertainties.
This determines $98$ values for $M_i^{ej}$ from $98$ data points and their $98$ uncertainties.
For completeness, we repeat the prior probabilities that are used here and elsewhere in the paper for the same parameters:
\begin{eqnarray}
	P(d_{known}) & \sim & \mathscr{N}_+(\mu_d,\sigma_d^2) \nonumber\\
	P(d_{unknown}) & \sim & \Gamma(\alpha_d, \beta_b) \nonumber\\
	P(\lg M^{ej}) & \sim & \mathscr{U}(\lg M_{min}^{ej},\lg M_{max}^{ej}) \nonumber \\
	P(R_N) & \sim & \mathscr{N}_+(\mu_{R_N},\sigma_{R_N}^2) \nonumber\\
	P(R_{ONe}) & \sim & \mathscr{N}_+(\mu_{R_{ONe}},\sigma_{R_{ONe}}^2) \label{eq:priors}
\end{eqnarray}
with $\mu_d$ and $\sigma_d$ from Tab.\,\ref{tab:results_table} in units of kpc, $\alpha_d = 4.25$ and $\beta_d = 1/2.45$ such that $\alpha_d/\beta_d = 10.4$\,kpc is the expectation value of CNe with unknown distances, $M_{min}^{ej} = 10^{-11}\,\mrm{M_{\odot}}$ and $M_{max}^{ej} = 10^{-4}\,\mrm{M_{\odot}}$, $\mu_{R_N} = 50\,\mrm{yr^{-1}}$ and $\sigma_{R_N} = 25\,\mrm{yr^{-1}}$, and $\mu_{R_{ONe}} = 16.7\,\mrm{yr^{-1}}$ and $\sigma_{R_{ONe}} = 8.3\,\mrm{yr^{-1}}$.
For the hierarchical model fits, instead of using truncated normal distributions for the event rates, we also test log-normal distributions with the same means and variances.
This avoids the unphysical boundary of a (nearly) zero rate.
The posterior distribution of the ejecta masses in each object, marginalised over the nuisance parameters distance and rate, finally reads
\begin{equation}
	P(M_i^{ej} | \langle F_i \rangle, \beta_i) \propto \iint \, dd_i \, dR \, P(\langle F_i \rangle\, | \,\beta_i, M_i^{ej}, d_i, R) P(M^{ej}, d_i, R)\mrm{.}
	\label{eq:full_posterior_no_pooling}
\end{equation}
In Eq.\,(\ref{eq:full_posterior_no_pooling}), $P(M^{ej}, d, R)$ is the joint prior distribution, built as multiplicative distribution from Eq.\,(\ref{eq:priors}).

We note that neither the true Poisson count data nor the condensed data set can constrain the distance (or rate) information which is consequently only a nuisance parameter and included to properly convert the flux values to ejecta masses, given the uncertainties in the distances (or rate).
Of our $97$ objects, certainly not all are ONe novae.
Nevertheless, we determine the upper bounds for the 1275\,keV line flux and the resulting \nuc{Na}{22} ejecta mass bounds from these objects.
A more detailed discussion the about the results and limits from individual objects is presented in Sec.\,\ref{sec:individual_object_results}.
The complete results table can be found in the Appendix, Tab.\,\ref{tab:results_table}.

\subsubsection{Complete Pooling}\label{sec:complete_pooling}
One way to improve upon estimates from individual objects -- in our case rarely being constraining (Sec.\,\ref{sec:individual_object_results}) -- is to include the whole population, and assume that each object (and the cumulative diffuse emission) stems from the same ejected mass.
Considering Fig.\,\ref{fig:model_graphical} this means that $M_i = \tilde{M}$, and the resulting fluxes $\langle F \rangle_i$ (and $\langle F \rangle_{diff}$) share a common dependency in the \textit{Complete Pooling} setting.
In lax terms, this is sometimes called `stacking' \citep[e.g.,][]{Malz2021_notstacking}, which however is not equivalent in detail because the extracted spectra (or fluxes) are \textit{not} stacked to obtain an average spectrum, but the model parameters are shared among each object.
For a completely pooled (combined) posterior for the ejected mass, we thus sample the fluxes according to the likelihood
\begin{equation}
	P( \langle F_i \rangle\, | \,\beta_i, \tilde{M}^{ej}, d_i, R_N) = \prod_{i=0}^{N_N+1} \Gamma(\langle F_i \rangle ; \tilde{F_i}\beta_i,\beta_i)\mrm{.}
	\label{eq:complete_pooling_sampling}
\end{equation}
The only difference between Eqs.\,(\ref{eq:complete_pooling_sampling}) and (\ref{eq:no_pooling_sampling}) is, that there is only one parameter of interest, $\tilde{M}^{ej}$, which is then representative for the whole population.
This determines $1$ value for $M^{ej}$ from $98$ data points and their $98$ uncertainties, dependent on each other, and still includes all uncertainties from distances, rates, and individual flux measurements.

\textit{Complete Pooling} results in the most optimistic however also most biased estimate of the ejecta masses.
Since the known effects of different ejecta masses for different WD masses and compositions are ignored in this setting, the resulting estimate is equivalent to the mean of the population of CNe.
This mean is not necessarily a useful number because it is dominated by the most abundant objects, most frequent CN types, or observations with the smallest uncertainties.

\subsubsection{Partial Pooling}\label{sec:partial_pooling}
Because of the different astrophysical and measurement-related effects, the true ejecta masses of CNe in the Milky Way may actually follow a distribution that can be described by a mean value $\mu$ and a width $\tau$, such that
\begin{equation}
	\lg M_i^{ej} \sim \mathscr{N}(\mu,\tau^2)\mrm{.}
	\label{eq:hyper_parameters}
\end{equation}
That means, we first sample ejecta masses according to the distribution, Eq.\,(\ref{eq:hyper_parameters}), and then again the flux distributions as before, but which now depend on the hyper-parameters $\mu$ and $\tau$ rather then $M_{i}^{ej}$ individually.
The likelihood becomes
\begin{equation}
	P( \langle F_i \rangle\, | \,\beta_i, \mu, \tau, d_i, R_N) = \prod_{i=0}^{N_N+1} \Gamma(\langle F_i \rangle ; \tilde{F_i}\beta_i,\beta_i)\mrm{.}
	\label{eq:partial_pooling_sampling}
\end{equation}
This \textit{Partial Pooling} setting is a compromise between \textit{No Pooling} and \textit{Complete Pooling}:
Depending on the value of $\tau$, the samples of  $\langle F_i \rangle$ either converge to one common value (\textit{Complete Pooling}, $\tau \rightarrow 0$) or spread out towards their independent values (\textit{No Pooling}, $\tau \rightarrow \infty$).
The prior for the mean of this hyper-distribution is chosen similar to the individual masses, $\mu \sim \mathscr{U}(-11,-4)$.
The prior for the width $\tau$ is discussed in the literature \citep[][, and references therein]{Gelman2013_BDA3}, depending on the purpose of the analysis.
We test two priors for $\tau$:
First, a uniform distribution between $0$ and $10$, which is agnostic about the fact that each object should indeed rather be similar in ejecta mass and should not show several orders of magnitude difference.
However, especially in the case of the 1275\,keV emission, such a prior would be useful because the ejecta masses would indeed be either close to zero (CO novae) or around a certain mean (ONe novae).
Second, we test a half-Cauchy prior for $\tau$ with width $1$.
This assumes that most objects follow a similar trend, but also allows for a broader distribution that might include several `outliers'.

We note that our data are mostly dominated by the measurement uncertainties so that the choice of the prior for $\tau$ has no influence on the result and both converge around the same point (Sec.\,\ref{sec:combined_analysis}).
\textit{Partial Pooling} determines the two shape parameters of the ejecta mass distribution, Eq.\,(\ref{eq:hyper_parameters}).
The upper bounds of $\mu$ can then be interpreted as conservative upper bounds on the ejecta masses of \nuc{Be}{7} and \nuc{Na}{22} in individual CN events in the Milky Way, taking into account the complete population, and its uncertainties from the SPI measurements.

\section{Results}\label{sec:results}

\subsection{Individual Objects}\label{sec:individual_object_results}
\begin{figure}[!hbtp]
	\centering
	\includegraphics[width=1.0\columnwidth,trim=0.4in 0.4in 0.8in 0.8in,clip=true]{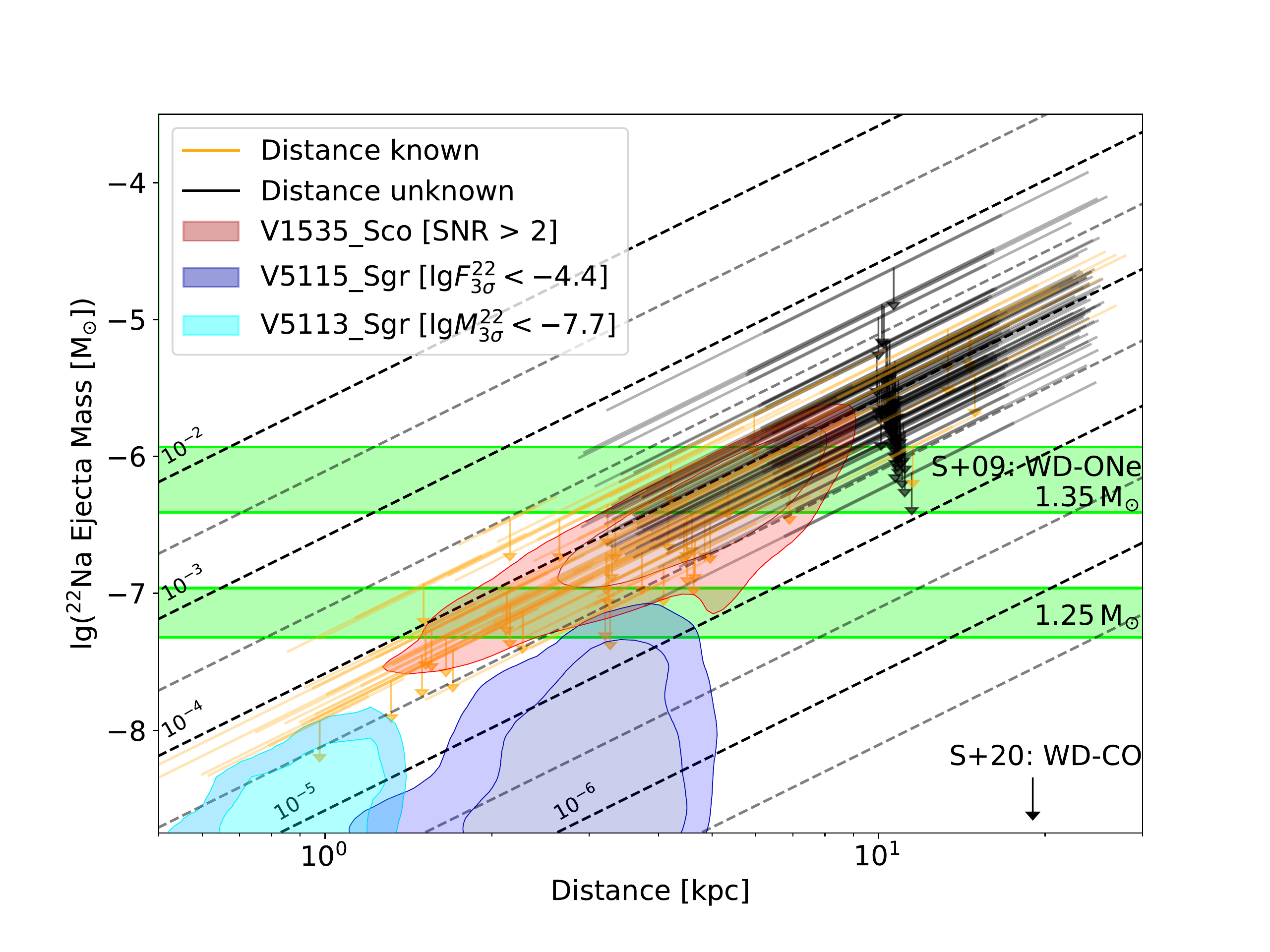}
	\caption{Summary of the \textit{No Pooling} analysis. Shown are the upper bounds (99.85th perc.) on the ejecta mass of \nuc{Na}{22} for each object with known (orange) and unknown (black) distance. The dependence on distance let the bounds on mass appear along the lines of constant flux (dashed lines). Full posteriors are given for objects with the best flux bound (V5115 Sgr), the best mass bound (V5113 Sgr), and whenever the significance is larger than $2\sigma$ (V1535 Sco). Theoretical expectations \citep{abcdefgh} are indicated by the green bands for $1.25$ and $1.35\,\mrm{M_{\odot}}$ ONe novae. Other theoretical expectations are bound to values below $\approx 10^{-8}\,\mrm{M_{\odot}}$ \citep[e.g.,][]{Jose1998_novae}. See text for discussion of individual objects.}
	\label{fig:no_pooling_results}
\end{figure}

In Fig.\,\ref{fig:no_pooling_results}, we illustrate the bounds on the \nuc{Na}{22} ejecta masses for each object at its (un)known distance, and how the estimates relate to upper bounds on the flux.
All our upper bounds are quoted on the 99.85th percentile level if not mentioned otherwise.
Apparently bright objects for which the flux is not consistent with zero within $2\sigma$, the full posterior distributions are shown (here only V1535 Sco).
Clearly, the mass estimates are inherently bound by the observation time
towards the target and the exponential decay law winning over the square-root of time increase in sensitivity.
Even though several objects are frequently re-visited since the beginning of the INTEGRAL mission, all become saturated in terms of significance above background.
This leads to an apparently common upper bound of the fluxes between $3 \times 10^{-5}$ and $3 \times 10^{-4}\,\mathrm{ph\,cm^{-2}\,s^{-1}}$ for \nuc{Na}{22} and $10^{-4}$ to $3 \times 10^{-3}\,\mathrm{ph\,cm^{-2}\,s^{-1}}$ for \nuc{Be}{7} (see Appendix Fig.\,\ref{fig:no_pooling_results_7Be}), which cannot be improved on the individual object basis.
Only a nearby ONe nova ($\lesssim 1.0\,\mrm{kpc}$) or one with exceptionally high ejecta mass would be seen with INTEGRAL/SPI.
The remaining variance in flux bounds and hence ejecta mass bounds stems from the varying background level in this 16 year data set, as well as exposure differences.

Nova V1535 Sco shows the largest significance above instrumental background with $2.5\sigma$.
With a distance of $5.3 \pm 1.7$\,kpc, this would be an unexpectedly large \nuc{Na}{22} ejecta mass.
Furthermore, it is not known whether V1535 Sco is an ONe or CO nova.
Hard X-ray emission ($> 1$\,keV) has been found in V1535 Sco within a few weeks after its discovery on Feb 11 2015, with a peculiar behaviour that is thought to be an indication of the CN being embedded in a red giant star \citep{Linford2017_V1535Sco}.
V1535 shows no to weak indications of GeV emission \citep{Franckowiak2018_FermiLATnovae}, probably because of its large distance.
We find this enhanced signal-to-noise ratio only for the 1275\,keV line, which would be unexpected if the emission was due to a CN outburst.

Among our sample, we obtain the lowest upper bound on the 1275\,keV line flux for V5115 Sgr.
The large exposure toward the Galactic centre (cf. Fig.\,\ref{fig:expo_map}) and its outburst time on Mar 28 2005 \citep{Nakano2005_V5115Sgr} leads to an upper bound on the 1275\,keV line flux of $<3.6 \times 10^{-5}\,\mrm{ph\,cm^{-2}\,s^{-1}}$.
V5115 Sgr is probably an ONe nova with a WD mass of $1.2 \pm 0.1\,\mrm{M_{\odot}}$ \citep{Hachisu2006_novae_declinelaw,Shara2018_WDmasses}.
At a distance of $2.8 \pm 1.0$\,kpc, the upper bound on the flux converts to an upper bound on the \nuc{Na}{22} ejecta mass of $<1.3 \times 10^{-7}\,\mrm{M_{\odot}}$.
This excludes WD masses around $1.3\,\mrm{M_{\odot}}$ (see Fig.\,\ref{fig:no_pooling_results}), fully consistent with theoretical expectations \citep{abcdefgh} and independent measurements \citep{Hachisu2006_novae_declinelaw}.
Using this bound, we can set a lower bound on the distance of V5115 Sgr of $\gtrsim 2.0$\,kpc, as otherwise, SPI would have seen the 1275\,keV line.
We note again that longer exposures for this object will not result in any improvements on the flux or ejecta mass, because most \nuc{Na}{22} has already decayed.

The tightest limit on the \nuc{Na}{22} ejecta mass, we obtain -- naturally -- for the object closest to us, here V5113 Sgr with a distance estimate of $0.9 \pm 0.2$\,kpc.
At this distance, we infer a bound on the mass up to $<2 \times 10^{-8}\,\mrm{M_{\odot}}$.
If V5113 Sgr was an ONe nova, this would set tight constraints on the nucleosynthesis on such objects, as it would fall below the lowest theoretical estimates \citep[e.g.][]{abcdefgh}.
There are no indications that V5113 Sgr is indeed an ONe nova \citep{Tanaka2011_nova_rebrightening,Mroz2015_OGLEnovacat}, nor is the mass of the WD constrained.

We note that the mass estimates for \nuc{Be}{7} and 478\,keV are one to two orders of magnitude above theoretical expectations in most cases and we refer to Appendix Sec.\,\ref{sec:additional_figures_Be7} for these results.

\subsection{Diffuse Emission}\label{sec:diffuse_emission_results}
\begin{figure}[!hbtp]
	\centering
	\includegraphics[width=1.0\columnwidth,trim=0.5in 0.6in 1.0in 1.0in,clip=true]{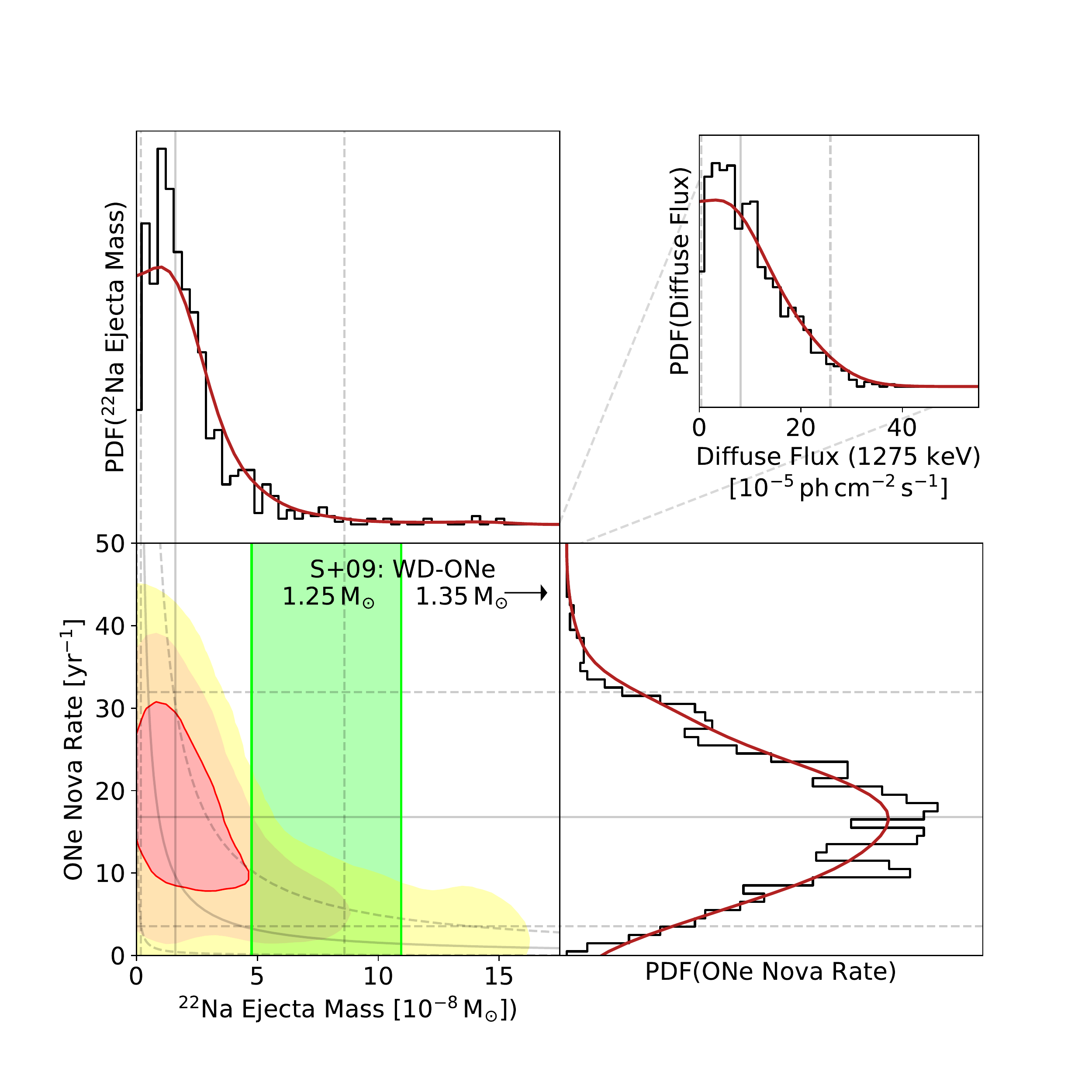}
	\caption{Posterior distributions for diffuse emission of the 1275\,keV line. The flux (\textit{top right}) is separated into its two determining components, the \nuc{Na}{22} ejecta mass (\textit{top left}) and the ONe nova rate (\textit{bottom right}, treated as a nuisance parameter). Theoretical predictions are indicated in green from \citet{abcdefgh}. The joint posterior (68.3th, 95.4th, 99.7th percentile) of ejecta mass and nova rate (\textit{bottom left}) shows the expected anticorrelated behaviour according to Eq.\,(\ref{eq:flux_diff_na22}).}
	\label{fig:diffuse_emission_posterior_22Na}
\end{figure}

In Fig.\,\ref{fig:diffuse_emission_posterior_22Na} we show the posterior distribution of the diffuse 1275\,keV line flux in the Galaxy, accounting for possible continuum emission, and separated into ONe nova rate and \nuc{Na}{22} ejecta mass.
The illustrated flux posterior takes into account that there is a weak diffuse Galactic continuum emission present in the band 1265--1285\,keV.
We determine the total flux in this band according to $F_{tot}^{22} = F_{Na}^{22} + F_{conti}^{22}$, where $F_{conti}^{22} \sim \mathscr{N}(7.6,2.5^2)$ in units of $10^{-5}\,\mrm{ph\,cm^{-2}\,s^{-1}}$ \citep{Wang2020_Fe60}.
From this, we find an upper bound on the diffuse 1275\,keV line flux of $<3.9 \times 10^{-4}\,\mrm{ph\,cm^{-2}\,s^{-1}}$ in the entire Galaxy, which converts to an upper bound on the \nuc{Na}{22} ejecta mass of $<2.7 \times 10^{-7}\,\mrm{M_{\odot}}$.
We want to point out that this value is similar to the one found by \citet{Jean2001_ONenovae1275} with COMPTEL ($<3 \times 10^{-7}\,\mrm{M_{\odot}}$), which only focused on the inner Galactic ridge ($|l| < 30^{\circ}$).
Our analysis takes into account the uncertainties on the ONe nova rate, the asymmetric flux estimate from the SPI count data, as well as possible continuum contribution, whereas the analysis by \citet{Jean2001_ONenovae1275} considered different spatial distributions for CNe.

The energy band 474--482\,keV is detected with more than $11\sigma$ above instrumental background with a flux of $F_{tot}^7 = (1.2 \pm 0.1) \times 10^{-3}\,\mrm{ph\,cm^{-2}\,s^{-1}}$.
Most of this flux is originating in the three-photon decay of ortho-Positronium in the interstellar medium, plus small contributions from Inverse Compton scattering and unresolved high-energy sources, for an integrated continuum flux of $F_{conti}^7 \sim \mathscr{N}(99.2,11.7^2)$ \citep[e.g.][]{Strong2005_gammaconti,Bouchet2011_diffuseCR,Siegert2016_511}.
Taking this into account yields an upper bound on the diffuse 478\,keV line flux of $<5.9 \times 10^{-4}\,\mrm{ph\,cm^{-2}\,s^{-1}}$, which converts to a \nuc{Be}{7} ejecta mass of $<4.1 \times 10^{-7}\,\mrm{M_{\odot}}$.
Again, the \nuc{Be}{7} values are not constraining and we refer to Appendix Sec.\,\ref{sec:additional_figures_Be7} for the results of the 478\,keV line.

\subsection{Combined Analysis}\label{sec:combined_analysis}
Since none of the objects are individually detected, the hierarchical analysis mainly takes into account the asymmetric uncertainties on these flux measurements.
This allows us to define common (population) upper bounds on the ejecta mass, given the individual uncertainties from fits to the SPI raw count data.
\begin{figure}[!hbtp]
	\centering
	\includegraphics[width=1.0\columnwidth,trim=0.1in 0.3in 0.4in 0.6in,clip=true]{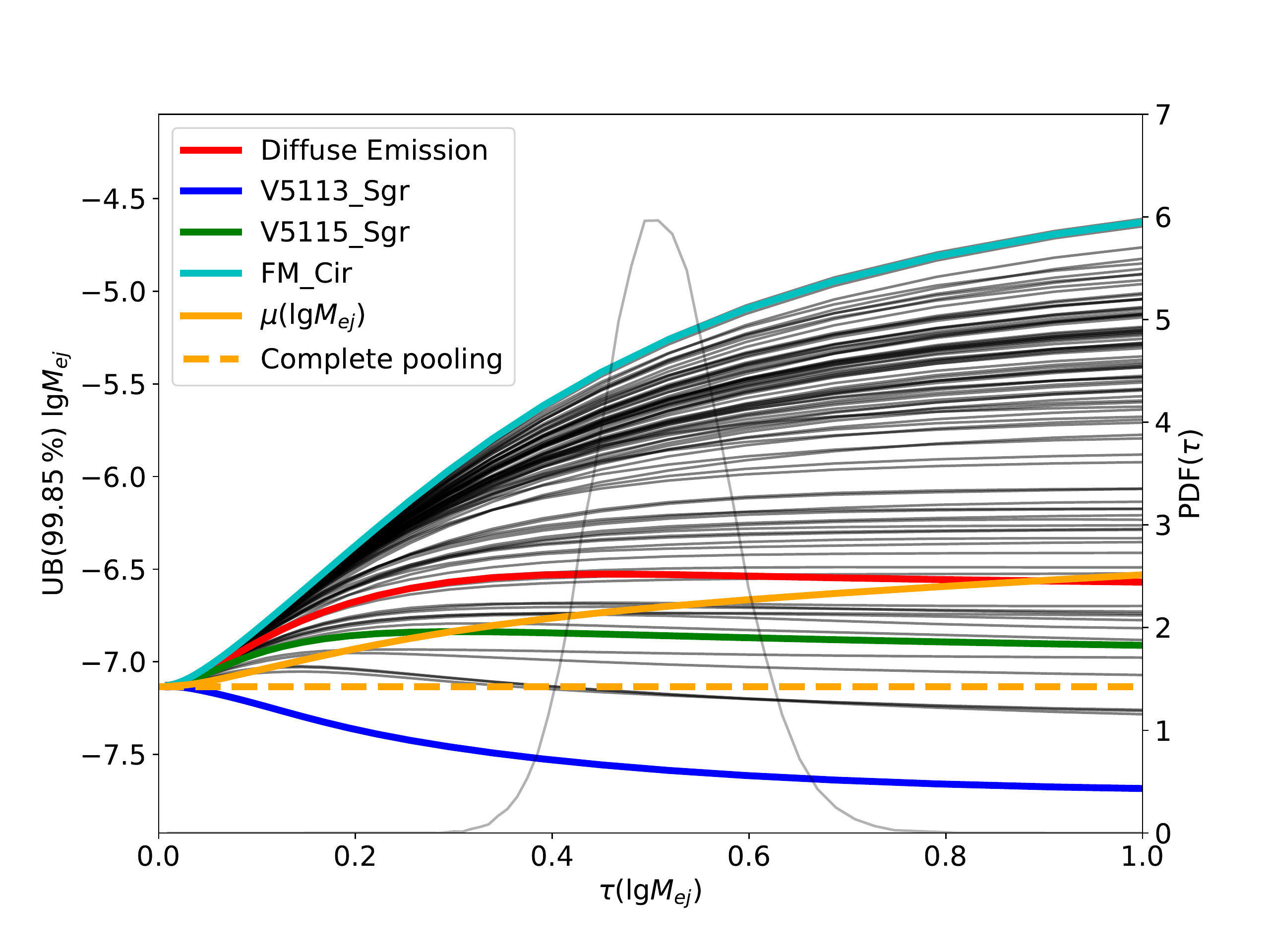}
	\caption{Summary of the hierarchical model. Shown are the upper bounds on the \nuc{Na}{22} ejecta masses (\textit{left} axis) for individual objects (black) as a function of the population width $\tau$, together with the \textit{Complete Pooling} estimate ($\tau \rightarrow 0$, dashed orange), and the population mean $\mu$ (solid orange). Individual objects of interest (best (blue) worst (aqua) bound, diffuse emission (red)) are indicated. As $\tau \rightarrow \infty$, each upper bound converges to its \textit{No Pooling} value (Tab.\,\ref{tab:results_table}). Given the SPI flux posteriors, a population width of $\tau = 0.52^{+0.18}_{0.23}$ (\textit{right} axis, gray shaded) is found. This value is dominated by the flux uncertainties.}
	\label{fig:summary_plot_hierarchical_model}
\end{figure}

In Fig.\,\ref{fig:summary_plot_hierarchical_model}, we show the summary of our hierarchical model for the \nuc{Na}{22} case.
As $\tau$ goes to zero in the case of \textit{Complete Pooling} (CP), all values (here: upper bounds on the ejecta mass) converge to one single point, $M_{ej,CP}^{22} < 0.7 \times 10^{-7}\,\mrm{M_{\odot}}$ (dashed orange line).
This assumes both, the ejecta mass of all individual point sources in our CN sample and the diffuse emission to originate from the same \nuc{Na}{22} ejecta mass.
We know that this is not the case (see Sec.\,\ref{sec:individual_object_results}), and we describe two ways of lifting these restrictions:

In an attempt to account for the fraction of $1/3$ of all CNe to be ONe novae, we use a finite mixture model with exactly this defined ratio $R_{CO} : R_{ONe} = 2 : 1$ and repeat the CP analysis.
This mixture model `picks' a sub-sample of one third of our sample and fits it according to Eq.\,(\ref{eq:complete_pooling_sampling}) whereas the remaining sample is subject to a prior distribution for the ejecta mass according to CO novae \citep[$\lg M_{22}^{ej,CO} \sim \mathscr{N}(-12,1^2)$,][]{Starrfield2020_COnovae}, resulting in basically zero flux for every object.
The fit results in an increased upper bound of $M_{ej,CP}^{22} < 1.0 \times 10^{-7}\,\mrm{M_{\odot}}$.

Alternatively, we use the \textit{Partial Pooling} approach (Sec.\,\ref{sec:partial_pooling}) and infer the upper bound on the population's ejecta mass by determining the spread $\tau$ of the distribution of ejecta masses.
Ideally $\tau$ would be determined by a combination of the true spread of the ejecta from many observations and their inherent uncertainties.
Here, $\tau$ is almost purely defined by the measurement uncertainties of the asymmetric flux posteriors.
We find that, independent of the choice on the prior of $\tau$ (cf. Sec.\,\ref{sec:partial_pooling}), its posterior populates the range $\tau = 0.52_{-0.23}^{+0.18}$ (C.I. 99.7\,\%) with a population mean $\mu_{22} < 2.0 \times 10^{-7}\,\mrm{M_{\odot}}$.
This means the `broadest allowed' ejecta mass distribution of the population of \nuc{Na}{22}-ejecting CNe that would still be consistent with our data is $\lg M_{22}^{ej} \sim \mathscr{N}(-6.7,0.7^2)$.
Or, in other words, if the mean of the ejecta mass distribution was any higher than $2 \times 10^{-7}\,\mrm{M_{\odot}}$, one or more objects (or the diffuse emission) would have been seen by SPI with at least $3\sigma$ above the instrumental background.
This value is larger than the tightest constraint from V5113 Sgr in the \textit{No Pooling} context ($M_{22}^{ej}(\mrm{V5113~Sgr}) < 1.8 \times 10^{-8}\,\mrm{M_{\odot}}$ as $\tau \rightarrow \infty$ and assuming V5113 Sgr is an ONe nova; cf. Tab.\,\ref{tab:results_table}), for example, however smaller than the one purely from diffuse emission.
This can be considered the most conservative upper bound on the \nuc{Na}{22} ejecta mass from ONe novae, because 1) it does not restrict high ejecta masses as would be case in the \textit{Complete Pooling} setting, 2) takes into account both, diffuse emission and known explosions, and 3) considers the entire Galaxy.
Using a log-normal prior for the rate decreases the upper bound on the population mean by less than 5\,\%.

While the rate has only a marginal impact on the final population estimate, the non-detection of individual sources also suppresses the diffuse emission flux, and hence lowers the ONe rate:
The initial prior mean was $\mu_{R_{ONe}} = 16.7\,\mrm{yr^{-1}}$ with a standard deviation of $\sigma_{R_{ONe}} = 8.3\,\mrm{yr^{-1}}$.
Now, the posterior mean of  $\mu_{R_{ONe}}$ is found at $6$ (\textit{Complete Pooling}) and $13\,\mrm{yr^{-1}}$ (\textit{Partial Pooling}), with a standard deviation of $3$ and $7\,\mrm{yr^{-1}}$, respectively, independent of the prior shape.
If all CNe happen to eject the same small amount of \nuc{Na}{22} as given in the \textit{Complete Pooling} setting, the ONe nova rate in the Milky Way would be at most $18\,\mrm{yr^{-1}}$ (99.85th percentile).
Clearly, such an assumption is extreme and the more conservative upper bound on the ONe nova rate from \textit{Partial Pooling} is $40\,\mrm{yr^{-1}}$, also independent of the prior shape.

In the case of \nuc{Be}{7} and the 478\,keV line, the shape of Fig.\,\ref{fig:summary_plot_hierarchical_model} is almost identical, only the numbers are spread further.
We find an upper bound on the \nuc{Be}{7} ejecta mass from \textit{Complete Pooling} of $M_{ej,CP}^{7} <0.5 \times 10^{-7}\,\mrm{M_{\odot}}$ and from \textit{Partial Pooling} of $\mu_7 < 2.5 \times 10^{-7}\,\mrm{M_{\odot}}$.
Both values are not constraining compared to theoretical expectations, which predict at most $4 \times 10^{-10}\,\mrm{M_{\odot}}$ \citep[e.g.][]{abcdefgh}.
However, individual CNe have been found with a larger \nuc{Be}{7} ejecta mass, on the order of several $10^{-9}\,\mrm{M_{\odot}}$ \citep[e.g.,][]{Molaro2016_V5668,Tajitsu2016_V5668,Tajitsu2015_novaBe7}, indeed suggesting a larger spread in ejecta masses than what is currently discussed.

\section{Discussion}\label{sec:discussion}

\subsection{Nova Nucleosynthesis}\label{sec:nova_nucleosynthesis}
All our upper bounds are consistent with theoretical expectations.
Considering the lowest theoretical expectations for $1.35\,\mrm{M_{\odot}}$ WDs from \citet{abcdefgh}, for example, of $\approx 3.9 \times 10^{-7}\,\mrm{M_{\odot}}$, we find 18 objects whose upper bounds on the ejecta mass would be below this threshold.
Among these 18 objects are V5115 Sgr and V5113 Sgr which we we already mentioned above (Sec.\,\ref{sec:individual_object_results}).
We find that $9$ objects are probably CO novae, $3$ are probably ONe novae, and $6$ are not determined.
Considering the three ONe novae V5115 Sgr, V382 Nor and V1187 Sco and our ejecta mass estimates ($M_{22}^{ej}(\mrm{V5115~Sgr}) < 1.3 \times 10^{-7}\,\mrm{M_{\odot}}$, $M_{22}^{ej}(\mrm{V382~Nor}) < 2.2 \times 10^{-7}\,\mrm{M_{\odot}}$, $M_{22}^{ej}(\mrm{V1187~Sco}) < 1.3 \times 10^{-7}\,\mrm{M_{\odot}}$; cf. Tab.\,\ref{tab:results_table}), we find that these events happened on WDs with masses less than $1.35\,\mrm{M_{\odot}}$.
This is fully consistent with the independent WD mass estimates from \citet{Shara2018_WDmasses}.
In general, our upper bounds can be interpreted that most ONe novae in the Milky Way occur on WDs with a mass of less than $1.35\,\mrm{M_{\odot}}$, as otherwise SPI would have seen a stronger signal.
This is expected because the mass distribution of measured WDs in the Galaxy is peaking around $1.13\,\mrm{M_{\odot}}$ with a width of $0.12\,\mrm{M_{\odot}}$ \citep{Shara2018_WDmasses}.
Other theoretical models predict lower \nuc{Na}{22} yields, on the order of $10^{-8}\,\mrm{M_{\odot}}$ \citep[e.g.,][]{Jose1998_novae}.
If these models are considered, our results cannot provide any constraints.

The upper bound on the \nuc{Na}{22} ejecta mass from ONe novae translates into a steady state \nuc{Na}{22} production of $3.3 \times 10^{-6}\,\mrm{M_{\odot}\,yr^{-1}}$.
On a time scale of 10\,Gyr, ONe novae hence produced at most $3.3 \times 10^4\,\mrm{M_{\odot}}$ of \nuc{Ne}{22}, assuming the ONe nova rate is constant.
With the solar abundance ratios by \citet{Lodders2003_abundances}, the Galaxy would contain a \nuc{Ne}{22} mass of $\approx 7.5 \times 10^4\,\mrm{M_{\odot}}$.
Clearly, a full population synthesis model would be required to compare the upper bound on the \nuc{Ne}{22} mass from CNe with the total Galactic content.
We nevertheless provide these order of magnitude evaluations to show a general broad consistency.

While the results from \nuc{Be}{7} are far from constraining, we can provide a similar upper bound on the total \nuc{Li}{7} mass inside the Milky Way.
With the upper bound on the \nuc{Be}{7} ejecta mass from the population of CNe of $\mu_{7}^{ej} < 2.5 \times 10^{-7}\,\mrm{M_{\odot}}$, we find that on a time scale of 10\,Gyr at most $1.25 \times 10^5\,\mrm{M_{\odot}}$ of \nuc{Li}{7} has been produced.
This is about 2--3 orders of magnitude above the theoretical values from \citet{Starrfield2020_COnovae}, who suggest that about $100\,\mrm{M_{\odot}}$ of the $1000\,\mrm{M_{\odot}}$ of \nuc{Li}{7} in the Milky Way is due to CO novae.
To further constrain the \nuc{Li}{7} production in the Milky Way from CNe with soft $\gamma$-rays, an instrument with a sensitivity improvement by two orders of magnitude would be required (see also next Sec.).

\subsection{Positrons from Classical Novae}\label{sec:positron_production}
Given the upper bound on the \nuc{Na}{22} ejecta mass from ONe novae in the Milky Way of $\mu_{22}^{ej} < 2.0 \times 10^{-7}\,\mrm{M_{\odot}}$, we can estimate how much CNe contribute at most to the Galactic positron budget.
With the probability of \nuc{Na}{22} to decay via $\beta^+$-decay of $p_{22}^+ = 0.904$, the production rate of positrons from the population of ONe novae is
\begin{equation}
	\dot{N}_{e^+} = R_{ONe} p_{22}^+ \frac{\mu_{22}^{ej}}{m_{22}} < 5.5 \times 10^{42}\,\mrm{e^+\,s^{-1}.}
	\label{eq:positron_production}
\end{equation}
\citet{Siegert2016_511} estimate a total positron annihilation rate in the Milky Way of $R_{e^+} \approx 5 \times 10^{43}\,\mrm{e^+\,s^{-1}}$.
Assuming a steady state annihilation in the Galaxy, this rate has to be sustained by the same production rate.
The population of ONe would therefore at most contribute to $10\,\%$ to the total Galactic positron production rate.
This is of similar magnitude as other known and validated source types, such as \nuc{Al}{26} from massive stars and their supernovae ($\approx 10\,\%$; $\tau_{26} = 1.05\,\mrm{Myr}$) and \nuc{Ti}{44} from core-collapse supernovae ($\approx 10\,\%$; $\tau_{44} = 86\,\mrm{yr}$).
For the remaining $80\,\%$ required for steady state annihilation, several other individual source types, populations, and scenarios have been suggested.
However until now, only indirect evidence for positron annihilation and production in these sources has been found \citep[cf.][for a review]{Prantzos2011_511}.

The population of CNe in the Galaxy will have a small but important contribution to the total positron production rate, complementing the different nucleosynthesis origins of Galactic positrons with frequent occurrences of CNe and comparably short lifetimes of parent nuclei.
These considerations do not exclude the very short-lived light $\beta^+$-unstable nuclei \nuc{N}{13} or \nuc{F}{18} to contribute significantly to the total content.
Because the expected very strong 511\,keV flash at the time of the explosion has never been observed, it might even be possible that these positrons also do not annihilate in the expanding nova cloud, but rather escape into the interstellar medium.
This would make CNe to a dominant source of positrons in the Milky Way.

Theoretical modelling of nucleosynthesis yields predict values close to our upper bounds, which suggests a bright future for the next generation of soft $\gamma$-ray telescopes, such as COSI \citep{Tomsick2019_COSI}, in both, nuclear decay as well as the 511\,keV positron annihilation line.

\section{Conclusion}\label{sec:conclusion}
In this study, we used 15 years of archival INTEGRAL/SPI $\gamma$-ray observations to infer nucleosynthesis ejecta masses of decaying \nuc{Na}{22} and \nuc{Be}{7} from the population of classical novae.
In our sample of 97 individual known objects, no signal is detected above a threshold of $3\sigma$.

The best upper bound for an individual CO nova is found for V5668 Sgr, owing to a dedicated observation campaign \citep[see][]{Siegert2018_V5668Sgr}, with a 478\,keV line flux of $<1.4 \times 10^{-4}\,\mrm{ph\,cm^{-2}\,s^{-1}}$, resulting in a \nuc{Be}{7} mass of $<3 \times 10^{-8}\,\mrm{M_{\odot}}$.
For the first time, we consider the diffuse 478\,keV line component of CO novae in the Milky Way, and find an upper flux bound of $<6.0 \times 10^{-4}\,\mrm{ph\,cm^{-2}\,s^{-1}}$, which converts to an average ejecta mass of $<4.1 \times 10^{-7}\,\mrm{M_{\odot}}$.
We improve these values by using a Bayesian hierarchical model of 478\,keV emitting novae, i.e. taking into account a common distribution of ejecta masses, and infer a population mean of $<0.5$--$2.5 \times 10^{-7}\,\mrm{M_{\odot}}$.
While these values are hardly constraining compared to theoretical expectations \citep[e.g.,][]{Starrfield2020_COnovae}, we nevertheless showed that proper modelling of sub-threshold sources can provide a significant improvement above the pure stacking of many point sources.

For individual ONe novae, our most constraining upper bounds on the \nuc{Na}{22} ejecta mass are found for V5115 Sgr and V1187 Sco with $<1.3 \times 10^{-7}\,\mrm{M_{\odot}}$.
The lowest 1275\,keV line upper bound is found for V2659 Cyg with $<2.9 \times 10^{-5}\,\mrm{ph\,cm^{-2}\,s^{-1}}$, however at an unknown distance.
The expected diffuse emission of the 1275\,keV line in the Milky Way along the Galactic plane with a strong peak in the bulge provides an upper flux bound of $<3.9 \times 10^{-4}\,\mrm{ph\,cm^{-2}\,s^{-1}}$, which converts to a \nuc{Na}{22} ejecta mass bound of $<2.7 \times 10^{-7}\,\mrm{M_{\odot}}$.
Our Bayesian hierarchical model for the 1275\,keV line results in a population mean of $<0.7$--$2.0 \times 10^{-7}\,\mrm{M_{\odot}}$.
These values are an improvement over previous studies:
1) the upper bounds on the ejecta masses are 2--5 (depending on the model assumptions) times smaller than from earlier COMPTEL measurements \citep{Jean2001_ONenovae1275}, and 2) they take into account both diffuse emission as well as known sources, their individual uncertainties, and uncertainties on circumstantial parameters, such as the nova rate and distances.

The result can be interpreted in a way that, if the population mean of ejecta masses was any higher, one or more ONe novae would have been detected with at least $3\sigma$ by INTEGRAL/SPI in this 15\,yr data set.
Still, these upper bounds are on the high side of theoretical expectations \citep[e.g.,][]{abcdefgh}:
we can exclude that most ONe novae happen on white dwarves with masses around $1.35\,\mrm{M_{\odot}}$, and thus finally probe the region of model calculations.
If we tighten the uncertainties on nova rate and individual distances, for example, and could at the same time make use of the full data set rather than relying on extracted fluxes, the hierarchical model would probably probe even deeper, i.e. setting tighter bounds on ejecta masses.

With \nuc{Na}{22} as a $\beta^+$-decayer, we find an upper bound on the positron production rate in the Milky Way from ONe novae of $< 5.5 \times 10^{42}\,\mrm{e^+\,s^{-1}}$.
Compared to the Galactic positron annihilation rate \citep{Siegert2016_511}, the population of ONe could make up to $10\,\%$ of the required positron production rate for a steady state configuration.

\section*{Software}
	\textit{OSA/spimodfit} \citep{spimodfit},
	\textit{numpy} \citep{Oliphant2006_numpy},
	\textit{matplotlib} \citep{Hunter2007_matplotlib},
	\textit{astropy} \citep{astropy2013_astropy},
	\textit{scipy} \citep{Virtanen2019_scipy},
	\textit{Stan/pystan} \citep{Carpenter2017_stan},
	\textit{arviz} \citep{Kumar2019_arviz}.

\begin{acknowledgements}
Thomas Siegert is supported by the German Research Foundation (DFG-Forschungsstipendium SI 2502/1-1 \& SI 2502/3-1). He thanks J. Michael Burgess for suggesting to do something important. This work used the Extreme Science and Engineering Discovery Environment (XSEDE), which is supported by National Science Foundation grant number ACI-1548562. We acknowledge the use of XSEDE resources via Startup allocation TG-PHY200045.
\end{acknowledgements}

\bibliographystyle{aa} 
\bibliography{new_bib} 

\appendix

\section{Additional Figure for \nuc{Be}{7}}\label{sec:additional_figures_Be7}
\begin{figure}[!htbp]
	\centering
	\includegraphics[width=1.0\columnwidth,trim=0.4in 0.4in 1.0in 0.9in,clip=true]{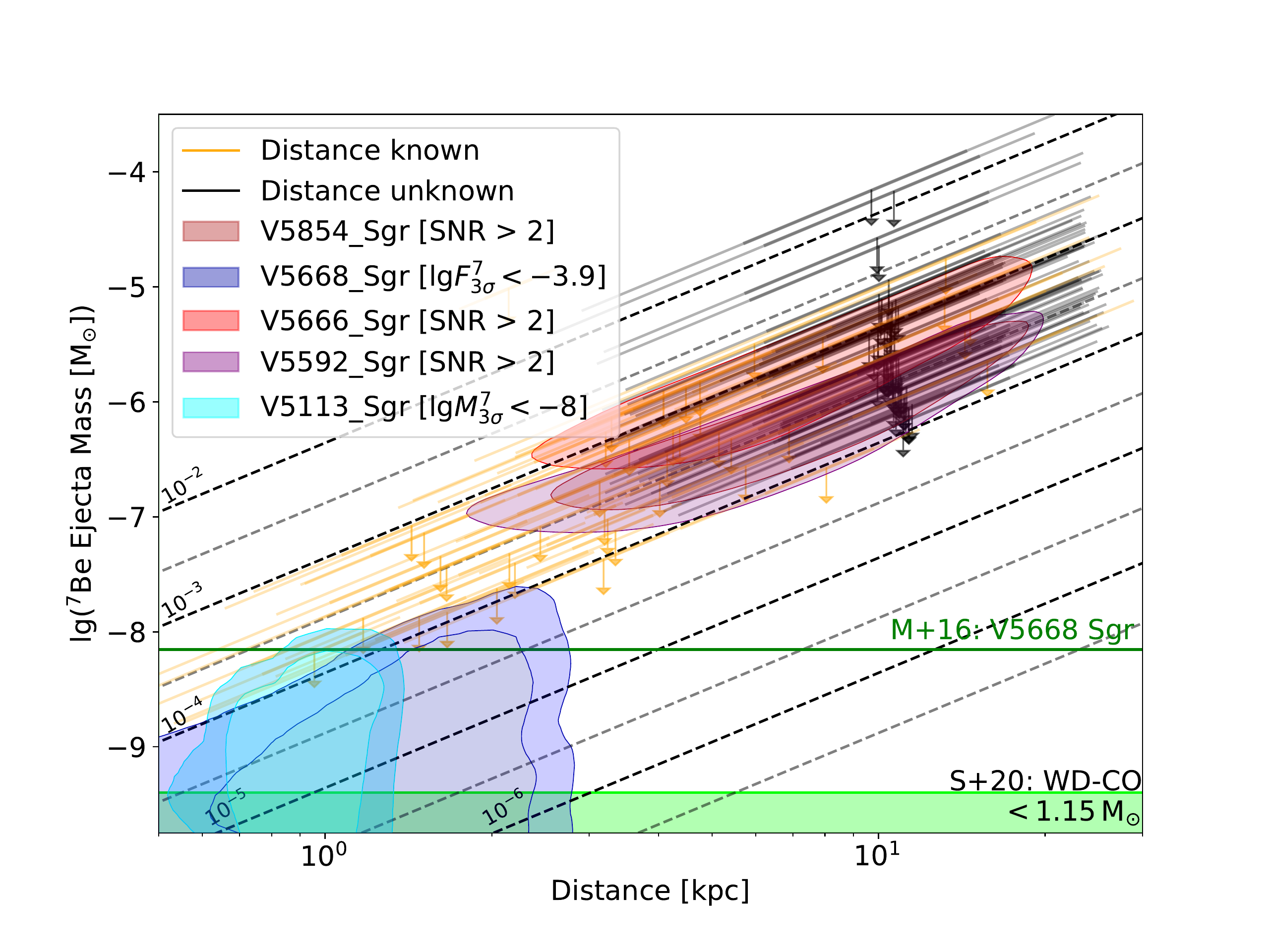}
	\caption{Same as Fig.\,\ref{fig:no_pooling_results} for the case of \nuc{Be}{7} and the 478\,keV line.}
	\label{fig:no_pooling_results_7Be}
\end{figure}

\begin{figure}[!hbtp]
	\centering
	\includegraphics[width=1.0\columnwidth,trim=0.5in 0.6in 1.0in 1.0in,clip=true]{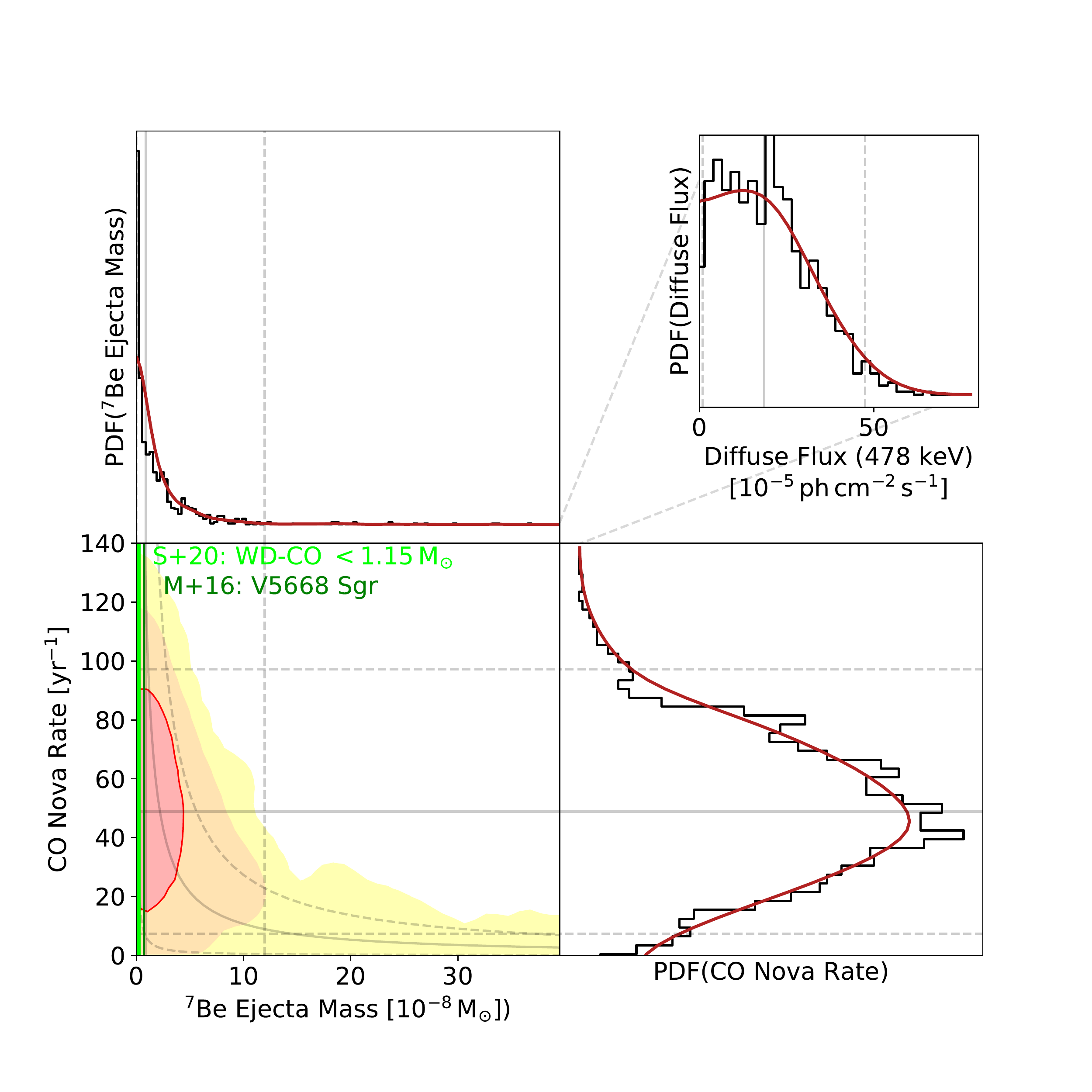}
	\caption{Same as Fig.\,\ref{fig:diffuse_emission_posterior_22Na} for the 478\,keV line.}
	\label{fig:diffuse_emission_posterior_7Be}
\end{figure}

\begin{figure}[!hbtp]
	\centering
	\includegraphics[width=1.0\columnwidth,trim=0.1in 0.3in 0.4in 0.6in,clip=true]{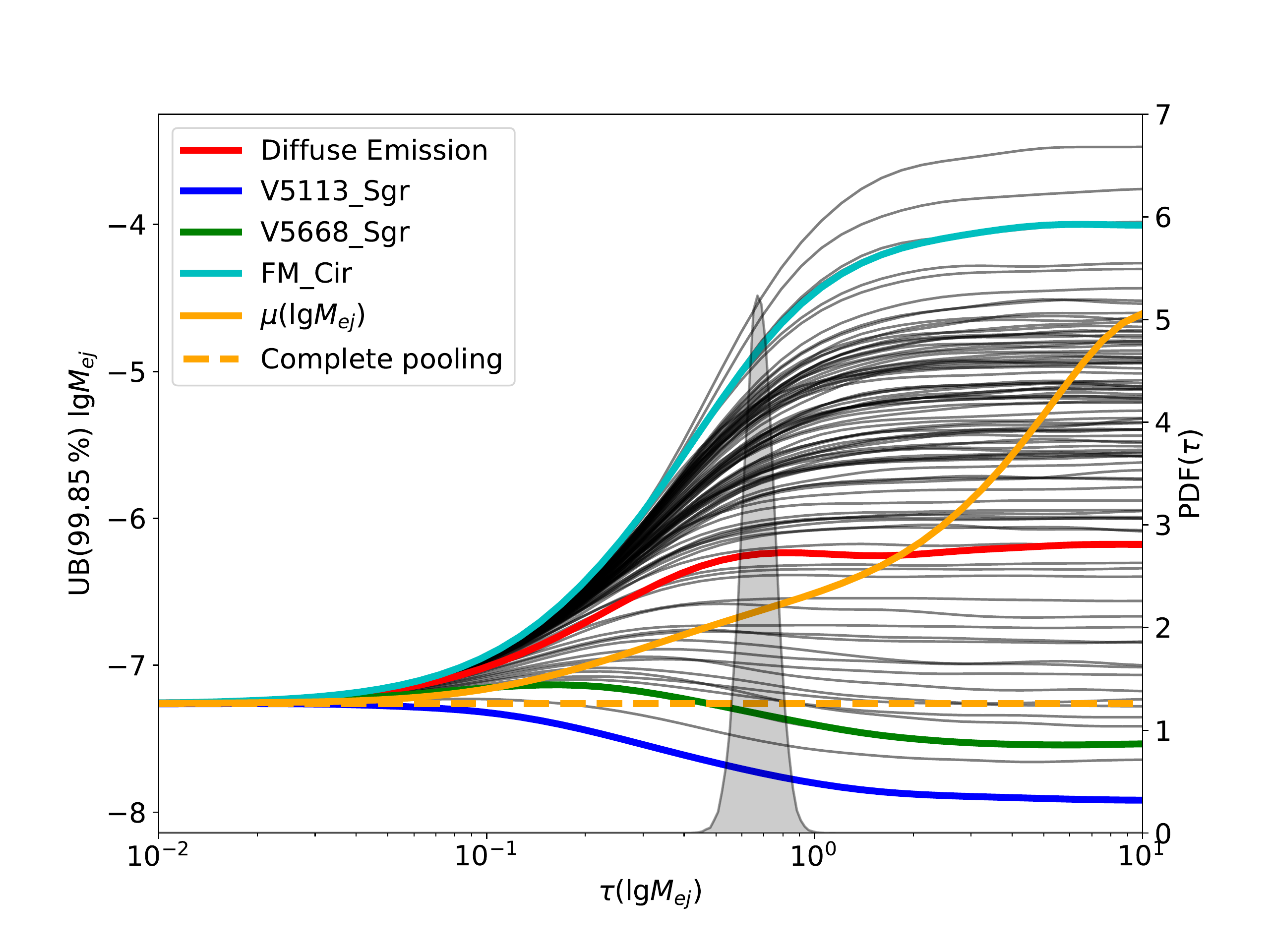}
	\caption{Same as Fig.\,\ref{fig:summary_plot_hierarchical_model} for \nuc{Be}{7}. The x-axis has been extended to show the convergence to the \textit{No Pooling} estimate for large $\tau$.}
	\label{fig:summary_plot_hierarchical_model_be7}
\end{figure}

The results for \nuc{Be}{7} from the measurements of the 478\,keV line are less constraining compared to \nuc{Na}{22} and the 1275\,keV line.
For completeness, we show the resulting upper bounds for \nuc{Be}{7} ejecta from individual objects in Fig.\,\ref{fig:no_pooling_results_7Be}, together with theoretical expectations \citep{Starrfield2020_COnovae} and one measured value for V5668 Sgr from \citet[][,M+16]{Molaro2016_V5668}.
In Fig.\,\ref{fig:diffuse_emission_posterior_7Be}, we show the joint posterior of the diffuse component of the 478\,keV line in the Milky Way, in which the strong continuum contribution is already marginalised out.
Finally, Fig.\,\ref{fig:summary_plot_hierarchical_model_be7} shows the summary of our hierarchical analysis in the case of \nuc{Be}{7}, with both extremes of \textit{Complete Pooling} ($\tau \rightarrow 0$) and \textit{No Pooling} ($\tau \rightarrow \infty$).
The posterior distribution of $\tau$ is completely determined by the uncertainties of the individual flux measurements.

\onecolumn

\section{Complete Results Table}\label{sec:results_table}
Our complete results are presented in Tab.\,\ref{tab:results_table}.

	\begin{longtable}{lrrrrrrrrr}
	\caption{ \label{tab:results_table} Summary table of results for individual objects, diffuse emission, and pooling analysis. From left to right, the columns are name of the object (or analysis type at the bottom), $T_0$ is the discovery date of the objects in units of $\mrm{MJD-50000}$, $l$ and $b$ are the Galactic longitude and latitude in degrees, $d$ and $\sigma_d$ are the distance and uncertainty estimates in units of kpc (see Sec.\,\ref{sec:nova_catalogue}). If no distance estimate is available, the generic distance prior (Sec.\,\ref{sec:unknown_distances}) has been used and marked as N/A in the table. $F_{22}^{ub}$ is the upper bound (99.85th percentile) of the 1275\,keV line flux, $M_{22}^{ub}$ the corresponding upper bound on the \nuc{Na}{22} ejecta mass, $F_{7}^{ub}$ is the upper bound on the 478\,keV line flux, and $M_{7}^{ub}$ is the corresponding bound on the ejected \nuc{Be}{7} mass. Fluxes are given in units of $10^{-5}\,\mrm{ph\,cm^{-2}\,s^{-1}}$, ejecta masses in units of $\mathrm{10^{-7}\,\mrm{M_{\odot}}}$.}\\
	\hline\hline
	Name &     $T_0$ &   $l$  &  $b$ &  $d$ &  $\sigma_d$ &  $F_{22}^{ub}$ &  $M_{22}^{ub}$  &     $F_{7}^{ub}$ &    $M_{7}^{ub}$ \\
	\hline
	\endfirsthead
	\caption{continued.}\\
	\hline\hline
	Name &     $T_0$ &   $l$  &  $b$ &  $d$ &  $\sigma_d$ &  $F_{22}^{ub}$ &  $M_{22}^{ub}$  &     $F_{7}^{ub}$ &   
	$M_{7}^{ub}$ \\
	\hline
	\endhead
	\hline
	\endfoot
	V2540 Oph &  2298.8 &    9.8 &   8.2 &   9.9 &       4.8 &   <6.6 &   <43.5 &  <2819.3 &  <963.3 \\
	V4743 Sgr &  2537.4 &   14.1 & -11.9 &   4.5 &       1.8 &   <6.2 &    <6.4 &   <178.8 &   <30.0 \\
	V4744 Sgr &  2572.4 &    4.9 &   2.1 &   N/A &       N/A &   <6.4 &   <67.1 &   <142.5 &  <149.2 \\
	DZ Cru &  2640.0 &  -60.5 &   2.3 &  10.7 &       2.9 &   <5.2 &    <4.0 &   <121.9 &   <98.1 \\
	V475 Sct &  2879.6 &   24.2 &  -3.9 &  14.0 &       6.7 &  <10.4 &  <197.9 &    <27.5 &   <58.1 \\
	V5113 Sgr &  2899.5 &    3.7 &  -4.1 &   1.0 &       0.2 &   <5.7 &    <0.2 &    <18.4 &    <0.1 \\
	DE Cir &  2921.0 &  -40.7 &  -3.8 &   N/A &       N/A &   <3.8 &   <24.3 &    <96.3 &  <131.8 \\
	V1186 Sco &  3189.1 &   -5.6 &   4.8 &   4.0 &       2.2 &   <6.1 &    <7.0 &    <52.6 &   <13.6 \\
	V1187 Sco &  3220.6 &   -4.3 &   1.5 &   3.2 &       0.1 &   <4.5 &    <1.3 &    <27.7 &    <1.3 \\
	V574 Pup &  3329.7 & -117.4 &  -2.0 &  12.4 &       5.9 &  <20.7 &  <253.4 &   <117.3 &  <225.8 \\
	V382 Nor &  3442.3 &  -27.7 &  -1.0 &   4.0 &       0.5 &   <4.3 &    <2.2 &    <31.8 &    <2.5 \\
	V5115 Sgr &  3457.8 &    6.0 &  -4.6 &   3.0 &       1.0 &   <3.6 &    <1.3 &    <15.9 &    <0.7 \\
	V5116 Sgr &  3555.0 &    2.1 &  -6.8 &   1.6 &       0.7 &   <7.4 &    <1.1 &    <41.4 &    <0.8 \\
	V1047 Cen &  3583.0 &  -53.7 &   0.0 &   2.9 &       2.0 &   <5.7 &    <3.2 &    <93.8 &   <11.1 \\
	V476 Sct &  3643.5 &   24.7 &   1.2 &  11.3 &       1.9 &   <4.3 &   <17.8 &    <19.5 &   <13.5 \\
	V5117 Sgr &  3783.4 &   -5.4 &  -6.4 &   1.4 &       0.3 &   <6.6 &    <0.5 &    <18.2 &    <0.2 \\
	V2576 Oph &  3831.6 &   -3.8 &   5.4 &   1.8 &       1.0 &   <5.9 &    <1.6 &    <17.1 &    <0.4 \\
	V1065 Cen &  4123.4 &  -66.0 &   3.6 &   3.3 &       0.5 &  <12.7 &    <4.8 &    <24.2 &    <1.3 \\
	V1280 Sco &  4135.9 &   -8.7 &   6.6 &   1.1 &       0.5 &   <6.2 &    <0.5 &    <24.4 &    <0.3 \\
	V2467 Cyg &  4174.8 &   80.1 &   1.8 &   1.5 &       0.3 &  <10.8 &    <1.0 &    <81.4 &    <1.0 \\
	V2615 Oph &  4178.8 &    4.1 &   3.3 &   2.1 &       0.8 &   <6.9 &    <1.6 &    <23.1 &    <0.7 \\
	V5558 Sgr &  4204.8 &   11.6 &   0.2 &   2.1 &       0.4 &   <9.0 &    <1.4 &    <30.9 &    <0.7 \\
	V458 Vul &  4316.5 &   58.6 &  -3.6 &   4.7 &       2.1 &  <11.8 &   <16.2 &    <57.2 &   <11.0 \\
	V598 Pup &  4320.0 & -110.9 & -13.8 &   2.1 &       0.2 &  <37.6 &    <5.0 &  <9177.2 &  <173.6 \\
	V459 Vul &  4459.3 &   58.2 &  -2.2 &  13.6 &       6.5 &  <13.4 &  <172.8 &    <90.1 &  <135.5 \\
	V2468 Cyg &  4532.8 &   66.8 &   0.2 &   6.8 &       1.0 &   <6.3 &    <9.7 &    <38.2 &    <8.8 \\
	NR TrA &  4557.7 &  -34.1 &  -7.2 &   4.5 &       1.6 &   <7.9 &    <6.3 &    <99.6 &   <13.8 \\
	V2491 Cyg &  4566.7 &   67.2 &   4.4 &   2.1 &       1.4 &  <14.6 &    <6.6 &    <35.3 &    <2.2 \\
	V5579 Sgr &  4574.8 &    3.7 &  -3.0 &  12.6 &       6.1 &  <10.5 &  <128.5 &   <175.6 &  <354.3 \\
	V1212 Cen &  4704.0 &  -46.1 &  -3.5 &   N/A &       N/A &  <15.9 &  <114.3 &    <88.9 &   <96.4 \\
	V1721 Aql &  4731.5 &   41.0 &  -0.1 &   7.5 &       2.0 &   <9.8 &   <21.0 &    <14.0 &    <5.3 \\
	QY Mus &  4738.0 &  -54.7 &  -4.9 &   5.6 &       2.4 &  <24.3 &   <45.9 &   <245.7 &   <75.1 \\
	V679 Car &  4796.3 &  -68.5 &  -0.5 &   4.5 &       2.3 &  <17.2 &   <16.5 &    <43.1 &    <9.4 \\
	V5580 Sgr &  4799.0 &    4.7 &  -6.5 &   N/A &       N/A &   <6.6 &   <44.5 &    <53.9 &   <74.3 \\
	V5582 Sgr &  4885.9 &    7.5 &   4.7 &   N/A &       N/A &   <8.2 &  <116.1 &    <19.5 &   <25.9 \\
	V5581 Sgr &  4942.7 &    2.2 &   1.8 &   N/A &       N/A &   <6.0 &   <45.7 &   <188.8 &  <206.5 \\
	V2672 Oph &  5059.5 &    1.0 &   2.5 &   3.1 &       0.7 &   <8.0 &    <2.8 &    <25.4 &    <1.4 \\
	V496 Sct &  5112.4 &   25.3 &  -1.8 &   3.2 &       0.8 &  <15.8 &    <6.9 &   <120.1 &   <10.5 \\
	V5585 Sgr &  5216.7 &    2.3 &  -4.2 &   N/A &       N/A &  <10.4 &  <158.4 &    <52.5 &   <51.9 \\
	V2674 Oph &  5245.8 &   -2.2 &   3.6 &   1.6 &       0.4 &   <7.3 &    <0.9 &    <36.5 &    <0.7 \\
	V1310 Sco &  5247.9 &  -11.5 &   2.2 &   N/A &       N/A &   <5.6 &   <28.2 &    <17.2 &   <24.3 \\
	V5586 Sgr &  5309.8 &    1.5 &  -1.0 &   N/A &       N/A &  <10.0 &  <126.6 &   <114.1 &  <145.9 \\
	V1723 Aql  &  5450.5 &   29.1 &  -0.9 &   N/A &       N/A &  <15.1 &  <151.2 &    <78.2 &  <156.2 \\
	V5587 Sgr &  5586.9 &    4.8 &   2.4 &   N/A &       N/A &   <8.3 &   <88.2 &    <53.0 &   <95.1 \\
	V5588 Sgr &  5647.8 &    7.8 &  -1.9 &   3.1 &       0.7 &   <7.2 &    <2.9 &    <50.7 &    <3.4 \\
	V1313 Sco &  5810.4 &  -18.4 &   3.8 &   N/A &       N/A &   <4.7 &   <38.3 &    <18.5 &   <24.6 \\
	V965 Per &  5872.0 &  151.6 & -17.9 &   N/A &       N/A &  <15.5 &   <85.1 &   <219.8 &  <325.0 \\
	V2676 Oph &  6011.8 &    0.3 &   5.3 &   N/A &       N/A &   <4.9 &   <26.6 &    <45.9 &   <76.4 \\
	V5589 Sgr &  6038.0 &    5.0 &   3.1 &   3.3 &       0.6 &   <8.9 &    <3.2 &   <147.7 &   <10.0 \\
	V1324 Sco &  6069.8 &   -2.6 &  -2.9 &   4.3 &       0.9 &   <5.3 &    <4.8 &    <76.1 &   <11.5 \\
	V5591 Sgr &  6104.5 &    7.2 &   2.5 &   3.6 &       2.3 &   <6.8 &    <8.9 &    <43.9 &    <8.3 \\
	V5592 Sgr &  6115.5 &    4.8 &  -6.1 &   N/A &       N/A &   <7.7 &   <99.8 &    <57.2 &  <170.7 \\
	V5593 Sgr &  6124.5 &   12.4 &  -1.9 &   N/A &       N/A &   <8.6 &  <104.7 &    <36.3 &   <70.5 \\
	V959 Mon &  6148.8 & -153.7 &   0.1 &   1.4 &       0.4 &  <17.9 &    <2.3 &   <168.5 &    <2.2 \\
	V1724 Aql &  6220.4 &   32.8 &  -0.4 &   N/A &       N/A &   <8.4 &   <39.4 &    <70.9 &  <152.1 \\
	V809 Cep &  6325.4 &  110.6 &   0.4 &   N/A &       N/A &   <6.5 &   <38.9 &   <180.5 &  <247.5 \\
	V1533 Sco &  6446.6 &   -7.4 &  -1.7 &   N/A &       N/A &   <7.3 &   <76.4 &    <81.7 &  <142.3 \\
	V339 Del &  6518.6 &   62.2 &  -9.4 &   3.6 &       2.2 &   <7.2 &    <5.6 &   <169.6 &   <24.6 \\
	V1830 Aql &  6593.5 &   37.1 &  -1.0 &   N/A &       N/A &  <13.4 &  <108.3 &  <1115.5 &  <978.7 \\
	V556 Ser &  6620.4 &   18.1 &   4.1 &   N/A &       N/A &   <9.5 &   <59.5 &   <367.2 &  <611.7 \\
	V5666 Sgr &  6683.9 &    9.9 &  -4.7 &   N/A &       N/A &   <7.3 &   <73.7 &   <168.6 &  <372.6 \\
	V2659 Cyg &  6747.8 &   70.5 &  -3.3 &   N/A &       N/A &   <2.9 &   <25.8 &    <35.1 &   <48.3 \\
	V1535 Sco &  7064.8 &  -10.1 &   3.9 &   5.3 &       1.7 &  <13.4 &   <22.7 &    <23.2 &    <6.4 \\
	V1658 Sco &  7068.0 &   -2.8 &  -2.4 &   N/A &       N/A &   <7.8 &   <67.2 &    <20.7 &   <31.4 \\
	V1404 Cen &  7081.3 &  -48.6 &  -1.6 &   N/A &       N/A &   <6.5 &   <54.3 &   <154.9 &  <183.9 \\
	V5668 Sgr &  7096.6 &    5.4 &  -9.9 &   1.4 &       0.7 &   <8.2 &    <0.8 &    <14.0 &    <0.3 \\
	V2944 Oph &  7110.8 &    6.6 &   8.6 &   N/A &       N/A &   <9.5 &   <62.2 &    <31.4 &   <54.1 \\
	V5669 Sgr &  7292.4 &    2.6 &  -3.1 &   N/A &       N/A &  <11.6 &  <123.3 &    <15.9 &   <21.7 \\
	V1831 Aql &  7300.5 &   49.8 &   0.3 &   N/A &       N/A &   <9.9 &   <77.9 &    <50.4 &   <55.7 \\
	V2949 Oph &  7306.4 &    2.8 &   4.6 &   N/A &       N/A &   <6.8 &   <50.2 &    <22.9 &   <36.2 \\
	V5850 Sgr &  7326.4 &   12.6 &  -2.6 &   N/A &       N/A &   <8.6 &   <57.4 &   <127.1 &  <195.4 \\
	V555 Nor &  7455.3 &  -33.1 &   1.6 &   N/A &       N/A &  <11.9 &  <159.9 &    <57.3 &   <44.1 \\
	V1655 Sco &  7549.6 &   -8.0 &  -3.2 &   N/A &       N/A &   <5.5 &   <41.6 &    <69.9 &  <138.5 \\
	V5854 Sgr  &  7581.2 &    0.2 &  -1.0 &   N/A &       N/A &  <10.3 &  <121.0 &    <53.6 &  <173.0 \\
	V5853 Sgr &  7608.5 &    3.8 &  -1.7 &   N/A &       N/A &   <4.4 &   <27.4 &    <31.2 &   <61.2 \\
	V1656 Sco &  7637.5 &   -5.2 &   2.5 &   N/A &       N/A &   <8.9 &   <82.9 &    <21.6 &   <36.4 \\
	V1659 Sco &  7638.0 &   -4.1 &  -1.9 &   N/A &       N/A &  <10.7 &  <145.0 &    <20.3 &   <33.6 \\
	V611 Sct &  7638.3 &   21.3 &   1.2 &   N/A &       N/A &  <15.8 &  <136.6 &    <34.1 &   <40.9 \\
	V407 Lup &  7655.0 &  -29.9 &   9.6 &   N/A &       N/A &  <18.4 &  <212.1 &    <90.5 &  <111.1 \\
	V5855 Sgr &  7681.4 &    4.0 &  -4.0 &  13.4 &       6.4 &   <7.5 &   <71.4 &    <43.0 &   <77.0 \\
	V5856 Sgr &  7686.0 &    4.3 &  -6.5 &   7.7 &       3.7 &   <7.9 &    <4.8 &    <91.6 &   <86.1 \\
	V1657 Sco &  7785.9 &  -13.7 &   3.9 &   N/A &       N/A &  <15.2 &  <169.8 &    <36.0 &   <71.2 \\
	V1405 Cen &  7868.1 &  -53.8 &  -1.0 &   N/A &       N/A &  <11.2 &   <89.8 &    <31.7 &   <42.6 \\
	V3662 Oph &  7881.8 &    2.7 &   3.2 &   N/A &       N/A &   <6.5 &   <54.6 &   <124.0 &  <199.4 \\
	V612 Sct &  7923.4 &   18.0 &  -2.2 &   4.0 &       1.0 &  <12.9 &    <7.9 &   <187.6 &   <22.0 \\
	V549 Vel &  8020.4 &  -92.8 &  -2.3 &   N/A &       N/A &  <36.8 &  <414.3 &   <114.5 &  <202.7 \\
	V1660 Sco &  8040.0 &   -3.6 &   1.6 &   N/A &       N/A &   <8.6 &   <46.8 &    <26.2 &   <33.9 \\
	V3663 Oph &  8068.4 &    0.1 &   7.2 &   N/A &       N/A &  <10.9 &   <65.5 &   <135.0 &  <234.8 \\
	V357 Mus &  8132.5 &  -65.7 &  -4.1 &   N/A &       N/A &  <23.4 &  <205.1 &    <56.2 &   <92.5 \\
	V1661 Sco &  8135.9 &   -5.9 &   3.2 &   N/A &       N/A &  <14.2 &  <114.7 &    <44.7 &   <67.4 \\
	FM Cir &  8137.7 &  -51.2 &  -5.3 &   N/A &       N/A &  <89.6 &  <497.7 &   <590.5 &  <713.8 \\
	V1662 Sco &  8155.9 &  -19.6 &  -0.1 &   N/A &       N/A &  <25.1 &  <248.5 &    <27.1 &   <34.2 \\
	V3664 Oph &  8161.8 &    1.3 &   6.4 &   N/A &       N/A &  <10.8 &   <66.6 &    <26.6 &   <46.0 \\
	V1663 Sco &  8173.4 &  -12.6 &   1.9 &   N/A &       N/A &  <11.9 &   <68.3 &    <28.8 &   <55.9 \\
	V3665 Oph &  8187.8 &   -3.7 &   5.8 &   N/A &       N/A &  <12.0 &   <91.4 &    <28.2 &   <46.5 \\
	V906 Car &  8197.3 &  -73.4 &  -1.1 &   N/A &       N/A &  <26.4 &  <230.6 &    <22.8 &   <18.9 \\
	V5857 Sgr &  8216.7 &   11.5 &   1.8 &   N/A &       N/A &  <39.6 &  <336.2 &    <44.0 &   <43.5 \\
	\hline
	Diffuse & - & - & - & - & - & <39.4 & <2.7 & <59.8 & <4.1 \\
	\hline
	\hline
	Complete Pooling & - & - & - & - & - & - & <0.7 (<1.0) & - & <0.5 \\
	Partial Pooling & - & - & - & - & - & - & <2.0 & - & <2.5 \\
	\hline
	\end{longtable}


\end{document}